\input harvmac

%
%
%
\message{S-Tables Macro v1.0, ACS, TAMU (RANHELP@VENUS.TAMU.EDU)}
%
%
\newhelp\stablestylehelp{You must choose a style between 0 and 3.}%
\newhelp\stablelinehelp{You should not use special hrules when
stretching
a table.}%
\newhelp\stablesmultiplehelp{You have tried to place an S-Table
inside another
S-Table.  I would recommend not going on.}%
%
%
\newdimen\stablesthinline
\stablesthinline=0.4pt
\newdimen\stablesthickline
\stablesthickline=1pt
%
%
\newif\ifstablesborderthin
\stablesborderthinfalse
\newif\ifstablesinternalthin
\stablesinternalthintrue
\newif\ifstablesomit
\newif\ifstablemode
\newif\ifstablesright
\stablesrightfalse
%
%
\newdimen\stablesbaselineskip
\newdimen\stableslineskip
\newdimen\stableslineskiplimit
%
%
\newcount\stablesmode
\newcount\stableslines
\newcount\stablestemp
\stablestemp=3
\newcount\stablescount
\stablescount=0
\newcount\stableslinet
\stableslinet=0
%
%
%
\newcount\stablestyle
\stablestyle=0
%
%
\def\stablesleft{\quad\hfil}%
\def\stablesright{\hfil\quad}%
%
%
\catcode`\|=\active%
%
%
\newcount\stablestrutsize
\newbox\stablestrutbox
\setbox\stablestrutbox=\hbox{\vrule height10pt depth5pt width0pt}
\def\stablestrut{\relax\ifmmode%
                         \copy\stablestrutbox%
                       \else%
                         \unhcopy\stablestrutbox%
                       \fi}%
%
%
\newdimen\stablesborderwidth
\newdimen\stablesinternalwidth
\newdimen\stablesdummy
\newcount\stablesdummyc
\newif\ifstablesin
\stablesinfalse
%
%
\def\begintable{\stablestart%
  \stablemodetrue%
  \stablesadj%
  \halign%
  \stablesdef}%
\def\stablesadj{%
  \ifcase\stablestyle%
    \hbox to \hsize\bgroup\hss\vbox\bgroup%
  \or%
    \hbox to \hsize\bgroup\vbox\bgroup%
  \or%
    \hbox to \hsize\bgroup\hss\vbox\bgroup%
  \or%
    \hbox\bgroup\vbox\bgroup%
  \else%
    \errhelp=\stablestylehelp%
    \errmessage{Invalid style selected, using default}%
    \hbox to \hsize\bgroup\hss\vbox\bgroup%
  \fi}%
\def\stablesend{\egroup%
  \ifcase\stablestyle%
    \hss\egroup%
  \or%
    \hss\egroup%
  \or%
    \egroup%
  \or%
    \egroup%
  \else%
    \hss\egroup%
  \fi}%
\def\stablestart{%
  \ifstablesin%
    \errhelp=\stablesmultiplehelp%
    \errmessage{An S-Table cannot be placed within an S-Table!}%
  \fi
  \global\stablesintrue%
  \global\advance\stablescount by 1%
  \message{<S-Tables Generating Table \number\stablescount}%
  \begingroup%
  \stablestrutsize=\ht\stablestrutbox%
  \advance\stablestrutsize by \dp\stablestrutbox%
  \ifstablesborderthin%
    \stablesborderwidth=\stablesthinline%
  \else%
    \stablesborderwidth=\stablesthickline%
  \fi%
  \ifstablesinternalthin%
    \stablesinternalwidth=\stablesthinline%
  \else%
    \stablesinternalwidth=\stablesthickline%
  \fi%
  \tabskip=0pt%
  \stablesbaselineskip=\baselineskip%
  \stableslineskip=\lineskip%
  \stableslineskiplimit=\lineskiplimit%
  \offinterlineskip%
  \def\borderrule{\vrule width \stablesborderwidth}%
  \def\internalrule{\vrule width \stablesinternalwidth}%
  \def\thinline{\noalign{\hrule height \stablesthinline}}%
  \def\thickline{\noalign{\hrule height \stablesthickline}}%
  \def\trule{\omit\leaders\hrule height \stablesthinline\hfill}%
  \def\ttrule{\omit\leaders\hrule height \stablesthickline\hfill}%
  \def\tttrule##1{\omit\leaders\hrule height ##1\hfill}%
  \def\stablesel{&\omit\global\stablesmode=0%
    \global\advance\stableslines by 1\borderrule\hfil\cr}%
  \def\el{\stablesel&}%
  \def\elt{\stablesel\thinline&}%
  \def\eltt{\stablesel\thickline&}%
  \def\elttt##1{\stablesel\noalign{\hrule height ##1}&}%
  \def\elspec{&\omit\hfil\borderrule\cr\omit\borderrule&%
              \ifstablemode%
              \else%
                \errhelp=\stablelinehelp%
                \errmessage{Special ruling will not display
properly}%
              \fi}%
  \def\stmultispan##1{\mscount=##1 \loop\ifnum\mscount>3
\stspan\repeat}%
  \def\stspan{\span\omit \advance\mscount by -1}%
  \def\multicolumn##1{\omit\multiply\stablestemp by ##1%
     \stmultispan{\stablestemp}%
     \advance\stablesmode by ##1%
     \advance\stablesmode by -1%
     \stablestemp=3}%

\def\multirow##1{\stablesdummyc=##1\parindent=0pt\setbox0\hbox\bgroup%

    \aftergroup\emultirow\let\temp=}
  \def\emultirow{\setbox1\vbox to\stablesdummyc\stablestrutsize%
    {\hsize\wd0\vfil\box0\vfil}%
    \ht1=\ht\stablestrutbox%
    \dp1=\dp\stablestrutbox%
    \box1}%
  \def\stpar##1{\vtop\bgroup\hsize ##1%
     \baselineskip=\stablesbaselineskip%
     \lineskip=\stableslineskip%

\lineskiplimit=\stableslineskiplimit\bgroup\aftergroup\estpar\let\temp
=}%
  \def\estpar{\vskip 6pt\egroup}%
  \def\stparrow##1##2{\stablesdummy=##2%
     \setbox0=\vtop to ##1\stablestrutsize\bgroup%
     \hsize\stablesdummy%
     \baselineskip=\stablesbaselineskip%
     \lineskip=\stableslineskip%
     \lineskiplimit=\stableslineskiplimit%
     \bgroup\vfil\aftergroup\estparrow%
     \let\temp=}%
  \def\estparrow{\vfil\egroup%
     \ht0=\ht\stablestrutbox%
     \dp0=\dp\stablestrutbox%
     \wd0=\stablesdummy%
     \box0}%
  \def|{\global\advance\stablesmode by 1&&&}%
  \def\|{\global\advance\stablesmode by 1&\omit\vrule width 0pt%
         \hfil&&}%
  \def\vt{\global\advance\stablesmode by 1&\omit\vrule width
\stablesthinline%
          \hfil&&}%
  \def\vtt{\global\advance\stablesmode by 1&\omit\vrule width
\stablesthickline%
          \hfil&&}%
  \def\vttt##1{\global\advance\stablesmode by 1&\omit\vrule width
##1%
          \hfil&&}%
  \def\vtr{\global\advance\stablesmode by 1&\omit\hfil\vrule width%
           \stablesthinline&&}%
  \def\vttr{\global\advance\stablesmode by 1&\omit\hfil\vrule width%
            \stablesthickline&&}%
  \def\vtttr##1{\global\advance\stablesmode by 1&\omit\hfil\vrule
width ##1&&}%
  \stableslines=0%
  \stablesomitfalse}
\def\stablesdef{\bgroup\stablestrut\borderrule##\tabskip=0pt plus
1fil%
  &\stablesleft##\stablesright%

&##\ifstablesright\hfill\fi\internalrule\ifstablesright\else\hfill\fi%

  \tabskip 0pt&&##\hfil\tabskip=0pt plus 1fil%
  &\stablesleft##\stablesright%

&##\ifstablesright\hfill\fi\internalrule\ifstablesright\else\hfill\fi%

  \tabskip=0pt\cr%
  \ifstablesborderthin%
    \thinline%
  \else%
    \thickline%
  \fi&%
}%
\def\endtable{\advance\stableslines by 1\advance\stablesmode by 1%
   \message{- Rows: \number\stableslines, Columns:
\number\stablesmode>}%
   \stablesel%
   \ifstablesborderthin%
     \thinline%
   \else%
     \thickline%
   \fi%
   \egroup\stablesend%
\endgroup%
\global\stablesinfalse}
%
%


\def\figin{\epsfcheck\figin}\def\figins{\epsfcheck\figins}
\def\epsfcheck{\ifx\epsfbox\UnDeFiNeD
\message{(NO epsf.tex, FIGURES WILL BE IGNORED)}
\gdef\figin##1{\vskip2in}\gdef\figins##1{\hskip.5in}
\else\message{(FIGURES WILL BE INCLUDED)}%
\gdef\figin##1{##1}\gdef\figins##1{##1}\fi}
\def\DefWarn#1{}
\def\figinsert{\goodbreak\midinsert}
\def\ifig#1#2#3{\DefWarn#1\xdef#1{fig.~\the\figno}
\writedef{#1\leftbracket fig.\noexpand~\the\figno}%
\figinsert\figin{\centerline{#3}}\medskip\centerline{\vbox{\baselineskip12pt
\advance\hsize by -1truein\noindent\footnotefont{\bf Fig.~\the\figno:} #2}}
\bigskip\endinsert\global\advance\figno by1}

\def \CM{{\cal M}}
\def \CP{{\cal P}}
\def \p{\partial}
\def \ges{{\ \lower-1.2pt\vbox{\hbox{\rlap{$>$}\lower5pt\vbox{\hbox{$\sim$}}}}\
}}
\def \les{{\ \lower-1.2pt\vbox{\hbox{\rlap{$<$}\lower5pt\vbox{\hbox{$\sim$}}}}\
}
}
\def\inbar{\,\vrule height1.5ex width.4pt depth0pt}
\font\cmss=cmss10 \font\cmsss=cmss10 at 7pt
\def\IR{\relax{\rm I\kern-.18em R}}
\def\IZ{\relax\ifmmode\mathchoice
{\hbox{\cmss Z\kern-.4em Z}}{\hbox{\cmss Z\kern-.4em Z}}
{\lower.9pt\hbox{\cmsss Z\kern-.4em Z}}
{\lower1.2pt\hbox{\cmsss Z\kern-.4em Z}}\else{\cmss Z\kern-.4em Z}\fi}
\def\IC{\relax\hbox{$\inbar\kern-.3em{\rm C}$}}
\def\IGa{\relax\hbox{${\rm I}\kern-.18em\Gamma$}}
\def\IPi{\relax\hbox{${\rm I}\kern-.18em\Pi$}}
\def\ITh{\relax\hbox{$\inbar\kern-.3em\Theta$}}
\def\IOm{\relax\hbox{$\inbar\kern-3.00pt\Omega$}}
\def\CM {{\cal M}}
\def\CF {{\cal F}}

\def\CR {{\cal R}}

\def\Diff{{\rm Diff}}
\def\CP {{\cal P}}

\def\IF{\relax{\rm I\kern-.18em F}}
\def\bO{{\IO}}
\def\IO{\relax\hbox{$\inbar\kern-.3em{\rm O}$}}

\def\jb{{\bar{\jmath}} }
\def\zb{{\bar{z}} }
\def\xb{{\bar{x}} }
\def\pib{{\bar{\pi}} }
\def\psib{{\bar{\psi}} }

\def\pb{{\bar{\partial}} }
\def\p{\partial}

\def\bG{{\rm \bf G}}

\def\Ad{\dot{A}}
\def\Bd{\dot{B}}


\def\unlockat{\catcode`\@=11}
\def\lockat{\catcode`\@=12}

\unlockat

\def\newsec#1{\global\advance\secno by1\message{(\the\secno. #1)}
\global\subsecno=0\global\subsubsecno=0\eqnres@t\noindent
{\bf\the\secno. #1}
\writetoca{{\secsym} {#1}}\par\nobreak\medskip\nobreak}
\global\newcount\subsecno \global\subsecno=0
\def\subsec#1{\global\advance\subsecno by1\message{(\secsym\the\subsecno. #1)}
\ifnum\lastpenalty>9000\else\bigbreak\fi\global\subsubsecno=0
\noindent{\it\secsym\the\subsecno. #1}
\writetoca{\string\quad {\secsym\the\subsecno.} {#1}}
\par\nobreak\medskip\nobreak}
\global\newcount\subsubsecno \global\subsubsecno=0
\def\subsubsec#1{\global\advance\subsubsecno by1
\message{(\secsym\the\subsecno.\the\subsubsecno. #1)}
\ifnum\lastpenalty>9000\else\bigbreak\fi
\noindent\quad{\secsym\the\subsecno.\the\subsubsecno.}{#1}
\writetoca{\string\qquad{\secsym\the\subsecno.\the\subsubsecno.}{#1}}
\par\nobreak\medskip\nobreak}

\def\subsubseclab#1{\DefWarn#1\xdef #1{\noexpand\hyperref{}{subsubsection}%
{\secsym\the\subsecno.\the\subsubsecno}%
{\secsym\the\subsecno.\the\subsubsecno}}%
\writedef{#1\leftbracket#1}\wrlabeL{#1=#1}}
\lockat



\def\inbar{\,\vrule height1.5ex width.4pt depth0pt}

\def\bG{{\IG}}

\def\bO{{\IO}}

\def\bR{{\IR}}

\def\cok{{\rm cok}\,}

\def\cF{{\cal F}}
\def\CF {{\cal F}}

\def\CL {{\cal L}}

\def\CM {{\cal M}}

\def\CO {{\cal O}}

\def\CP {{\cal P }}
\def\CR {{\cal R}}
\def\CW{{\cal W}}

\def\CZ {{\cal Z}}
\def\det{{\rm det}}
\def\Det{{\rm Det}}

\def\O{\Omega}

\def\half{{\textstyle{1\over 2}}}

\def\IB{\relax{\rm I\kern-.18em B}}
\def\ib{{\bar \imath}}
\def\IC{\relax\hbox{$\inbar\kern-.3em{\rm C}$}}
\def\ID{\relax{\rm I\kern-.18em D}}
\def\IE{\relax{\rm I\kern-.18em E}}
\def\IF{\relax{\rm I\kern-.18em F}}
\def\IG{\relax\hbox{$\inbar\kern-.3em{\rm G}$}}
\def\IGa{\relax\hbox{${\rm I}\kern-.18em\Gamma$}}
\def\IH{\relax{\rm I\kern-.18em H}}
\def\II{\relax{\rm I\kern-.18em I}}
\def\IK{\relax{\rm I\kern-.18em K}}
\def\IL{\relax{\rm I\kern-.18em L}}
\def\IM{\relax{\rm I\kern-.18em M}}
\def\IN{\relax{\rm I\kern-.18em N}}
\def\IO{\relax\hbox{$\inbar\kern-.3em{\rm O}$}}
\def\Iom{{\inbar\kern-3.00pt\Omega}}
\def\IOm{\relax\hbox{$\inbar\kern-3.00pt\Omega$}}
\def\IP{\relax{\rm I\kern-.18em P}}
\def\IPi{\relax\hbox{${\rm I}\kern-.18em\Pi$}}

\def\IQ{\relax\hbox{$\inbar\kern-.3em{\rm Q}$}}
\def\IR{\relax{\rm I\kern-.18em R}}
\def\ITh{\relax\hbox{$\inbar\kern-.3em\Theta$}}

\font\cmss=cmss10 \font\cmsss=cmss10 at 7pt
\def\IZ{\relax\ifmmode\mathchoice
{\hbox{\cmss Z\kern-.4em Z}}{\hbox{\cmss Z\kern-.4em Z}}
{\lower.9pt\hbox{\cmsss Z\kern-.4em Z}}
{\lower1.2pt\hbox{\cmsss Z\kern-.4em Z}}\else{\cmss Z\kern-.4em
Z}\fi}

\def\log {{\rm log}}

\def\p {\partial}
\def\pb{\bar{\partial}}

\def\zb {{\bar{z}}}
%
%

\def\dim{\mathop{\rm dim}}

\def\np{\nabla_+}

\def\lieg{{
{\bf g}}}


\def\boxit#1{\vbox{\hrule\hbox{\vrule\kern8pt
\vbox{\hbox{\kern8pt}\hbox{\vbox{#1}}\hbox{\kern8pt}}
\kern8pt\vrule}\hrule}}
\def\mathboxit#1{\vbox{\hrule\hbox{\vrule\kern8pt\vbox{\kern8pt
\hbox{$\displaystyle #1$}\kern8pt}\kern8pt\vrule}\hrule}}

\def\Xhat{{\widehat{X}}}
\def\Xhhat{{\widehat{\widehat{X}}}}
\def\itE{{\widehat{\widehat{E}}}}
\def\Ehat{{\widehat{E}}}

\def\uhat{{\widehat{u}}}
\def\psihat{{\widehat{\psi}}}

\def\ext{{\raise.3ex\hbox{$\textstyle \bigwedge$}}}
\def\e{\epsilon}
\def\Vhat{\widehat{V}}
\def\Vhhat{\widehat{\widehat{V}}}
\def\uhat{{\widehat{u}}}
\def\psihat{{\widehat{\psi}}}
\def\Hhat{{\widehat{H}}}
\def\Uhat{{\widehat{U}}}



\lref\barret{J. Barrett, G.W. Gibbons, M.J. Perry, C.N.
Pope and P.J.  Ruback, ``Kleinian geometry and the N=2 string,''
 Int. J. Mod. Phys. {\bf A9} (1994) 1457-1494, hep-th/9302073.}

\lref\brkvf{N. Berkovits and C. Vafa, ``N=4 Topological Strings,''
Nucl. Phys. {\bf B433}(1995) 123--180, hep-th/9407190.}

\lref\bgv{N.~ Berline, E.~ Getzler, and M.~ Vergne, {\it Heat
Kernels and Dirac Operators}, Springer 1992}

\lref\calabi{E. Calabi, Ann. Sci. Ecole Norm. Sup.
{\bf 12} (1979) 269; See also, S. Cecotti,
S. Ferrara, and L. Girardello,`` Geometry
of Type II Superstrings and the moduli of
superconformal field theories,'' Int. J. Mod. Phys.
{\bf A4} (1989) 2475}

\lref\CMRII{S.~ Cordes, G.~ Moore, and S.~ Ramgoolam,
``Large N 2D Yang-Mills Theory and Topological
String Theory'' hep-th/9402107. To appear in  Commun.  in Math. Phys}

\lref\CMRIII{S.~ Cordes, G.~ Moore, and S.~ Ramgoolam,
in Les Houches Session LXII on http://xxx.lanl.gov/lh94.}

\lref\nwtw{R. Dijkgraaf,
K. Intriligator, G. Moore, R. Plesser,
``A new twist on topological sigma models,'' unpublished.}

\lref\Kal{J.~ Kalkman, ``BRST Model for Equivariant Cohomology
and Representatives for the Equivariant Thom Class", Commun. Math.
Phys. {\bf 153} (1993) 447; ``BRST Model Applied to Symplectic Geometry",
PRINT-93-0637 (Utrecht), hep-th/9308132.}

\lref\mooreicm{G. Moore, ``2D Yang-Mills Theory and
Topological Field Theory,'' hep-th/9409044, Proceedings of
ICM94}

\lref\ogvf{H. Ooguri and C. Vafa,
``Geometry of $N=2$ Strings,''
Nucl. Phys. {\bf B361}  (1991) 469-518.}

\lref\ovii{H. Ooguri and C. Vafa, ``All Loop N=2
String Amplitudes,'' Nucl. Phys. {\bf B451} (1995) 121--161,
hep-th/9505183.}

\lref\park{J.-S. Park, ``Holomorphic Yang-Mills theory on compact Kahler
manifolds,'' hep-th/9305095; Nucl. Phys. {\bf B423} (1994) 559;
J.-S.~ Park, ``$N=2$ Topological Yang-Mills Theory on Compact K\"ahler
Surfaces", Commun. Math, Phys. {\bf 163} (1994) 113;
J.-S.~ Park, ``$N=2$ Topological Yang-Mills Theories and Donaldson
Polynomials", hep-th/9404009.}

\lref\VaWi{C.~ Vafa and E.~ Witten, ``A Strong Coupling Test of $S$-Duality",
 Nucl. Phys. {\bf B431} (1994) 3--77, hep-th/9408074.}

\lref\trminus{E.~ Witten, ``Constraints on Supersymmetry  Breaking",
Nucl. Phys. {\bf B202} (1982) 253.}

\lref\wttngrv{E. Witten,
``Topology Changing Amplitudes in
$(2+1)$-Dimensional Gravity,''
Nucl. Phys. {\bf B323} (1989) 113.}

\lref\gross{D. Gross and W. Taylor, ``Two Dimensional QCD
is a String Theory,'' Nucl. Phys. {\bf B400} (1993) 181--210,
hep-th/9301068.}

\lref\floer{A. Floer, ``Symplectic Fixed Points and Holomorphic
Spheres,'' Commun. Math. Phys. {\bf 120} (1989) 575.}

\lref\blau{M.~Blau and G.~Thompson, ``$N=2$
topological gauge theory, the Euler characteristic of moduli spaces,
the Casson invariant,'' Commun. Math. Phys. {\bf 152} (1993) 41-72.}

\lref\blauthom{M.~Blau and G.~Thompson, 
``Topological Gauge Theories from Supersymmetric Quantum Mechanics on
Spaces of Connections,'' Int.  Journal Mod. Phys. {\bf A8} (1993)
573--585.}

\lref\Witdgtr{E.~ Witten, ``Two Dimensional Gauge Theories Revisited",
J. Geom. Phys. {\bf G9} (1992) 303; hep-th/9204083.}

\lref\Witr{E.~ Witten, ``Introduction to Cohomological Field Theories",
Lectures at Workshop on Topological Methods in Physics, Trieste, Italy,
Jun 11-25, 1990, Int. J. Mod. Phys. {\bf A6} (1991) 2775.}

\lref\wittenmirror{E.\ Witten, in
{\sl Essays on Mirror manifolds,} Ed. S-T Yau (International Press,
Hong Kong, 1992).}

\lref\wittensigma{E.\ Witten,  
``Topological sigma models,'' Commun.  Math. Phys.  {\bf 118} (1988)
 411; ``On the topological phase of two dimensional gravity,''
 Nucl. Phys. {\bf B340} (1990) 281; `` Two dimensional gravity and
 intersection theory on moduli space,'' Surveys In Diff.  Geom. {\bf
 1} (1991) 243.}

\lref\polyakov{A.M~ Polyakov, ``Fine Structure of Strings",
Nucl. Phys. {\bf B268} (1986) 406.}

\lref\horava{ P. Horava, ``Topological Strings and QCD in Two Dimensions,''
hep-th/9311156; ``Topological Rigid String Theory and Two Dimensional QCD,''
hep-th/9507060.}

\lref\tftalso{M. Blau, G. Thompson,
``Localization and Diagonalization: A review of functional integral 
techniques for low-dimensional gauge theories and topological 
field theories,''  J. Math. Phys. {\bf 36} (1995) 2192--2236,
hep-th/9501075.}

\lref\Dubois{M. Dubois-Violette, ``A bigraded version of the Weil
alegbra and of the Weil homomorphism for Donaldson invariants,''
J. Geom. Phys. {\bf 19} (1996) 18--30, hep-th/9402063.}

\Title{ \vbox{\baselineskip12pt\hbox{hep-th/9608169}
\hbox{ YCTP-P17-96  }}}
{\vbox{
\centerline{Balanced Topological Field Theories }
}}
\bigskip
\centerline{Robbert Dijkgraaf}
\vskip.05in
\centerline{\sl Department of Mathematics }
\centerline{\sl University of Amsterdam }
\centerline{\sl 1018 TV Amsterdam }
\medskip
\centerline{and}
\medskip
\centerline{Gregory Moore}
\vskip.05in
\centerline{\sl Department of Physics}
\centerline{\sl Yale University}
\centerline{\sl New Haven, CT 06520}
\bigskip

We describe a class of topological field theories called ``balanced
topological field theories.''  These theories are associated to moduli
problems with vanishing virtual dimension and calculate the Euler
character of various moduli spaces.  We show that these theories are
closely related to the geometry and equivariant cohomology of
``iterated superspaces'' that carry two differentials. 
We find the most general action for
these theories, which turns out to define Morse theory on field
space. We illustrate the constructions with numerous examples.
Finally, we relate these theories to topological sigma-models
twisted using an isometry of the target space.

\Date{August  26, 1996}

\newsec{Introduction and Conclusion}

In recent years   several examples of topological
quantum field theories that compute the Euler number of particular
moduli spaces have been investigated.
For example, in
 \CMRII\CMRIII\ it was shown that the large $N$
expansion of two-dimensional Yang-Mills theory \gross\ has a natural
interpretation in terms of holomorphic maps from one Riemann surface
to another.  In order to write down a world-sheet action for this
string theory topological field theories were constructed that
calculate the Euler characters of moduli spaces of holomorphic maps. A
very similar construction was employed in \VaWi\ to explain the
occurrence of Euler characters of the moduli space of anti-selfdual
connections in the topologically twisted $N=4$ supersymmetric
Yang-Mills theory. The purpose of this paper is to clarify the
underlying geometry of the constructions of \CMRII\CMRIII\VaWi\ and to
generalize these to a class of topological field theories we call
``balanced topological field theories'' (BTFT's).

These models have the characteristic property of possessing two
topological charges $d_\pm$ and could very well be called $N_T=2$
topological field theories, where $N_T$ is the number of {\it topological}
charges. However, this might perhaps be confusing terminology, since
topological field theories with one topological charge ($N_T=1$) are
typically obtained by twisting  supersymmetric field theories with two
supercharges ($N=2$). Because of this possible confusion and because
the ghost numbers are perfectly matched in these models, we prefer to
use the term balanced field theories.

It turns out that these theories are intimately connected with a class
of superspaces we call iterated superspaces. These spaces carry two
exterior differentials $d_\pm$. We will show how the equivariant
cohomology of these superspaces leads naturally to the peculiar field
multiplets appearing in \CMRII\CMRIII\VaWi.  Moreover, these theories
have a fundamental $sl_2$ symmetry acting on the field space. The two
BRST symmetries $d_\pm$ transform covariantly under this $sl_2$
symmetry. Since the formalism of topological field theory is very
closely tied to de Rahm cohomology, fiber bundles and equivariant
cohomology, we will in this paper develop the generalizations of these
concepts to the extended case.

One of the simplifying properties of balanced field theories is that the
action can be determined from an {\it action
potential} $\cF$:
\eqn\actpoti{
S  = d_+ d_- \CF.
}
As we will explain, $\CF$ should be thought of as a Morse function on
field space. The path integral localizes to the critical points of
$\cF$. In our formalism gauging a symmetry is a trivial operation: One
simply uses the differentials of equivariant cohomology in \actpoti.
This aspect is but one of the various simplifying properties of
$N_T=2$ topological field theories.

In this paper we focus on the geometrical foundations of the theory,
and just briefly indicate the various applications.  The outline of
the paper is as follows. In section 2 we discuss the geometry of
balanced topological theories in terms of  iterated
superspaces.  We pay particular attention to the case where the
underlying bosonic manifold is the total space of a vector bundle.
Also, familiar concepts such
as the Lie derivative and the inner product are
given their appropriate generalizations. In section 3 we treat the
equivariant case. For extended topological symmetry the geometry of
principal bundles becomes very rich. In particular we will see that
the curvature gets replaced by a full multiplet, consisting of
a triplet of bosonic 2-form curvatures together with a doublet of
fermionic 3-form curvatures. In section 4 we formulate balanced topological
theories using the geometrical formalism developed in sections 2 and
3. We prove the existence of an action potential and make contact with
the co-field formalism of \CMRII\CMRIII\ .  In section 5 we prove the
localization properties of a BTFT and show that it computes the
Euler number.  Also a very elegant formalism of
gauge fixing is mentioned.  Section 6 points out various examples,
but they are not all treated in depth. Finally, section 7 contains a
discussion of the relation with topological sigma-models, using a
so-called isometry twist.

Finally we would like to point out that some aspects of our
construction relating topological $N_T=2$ theories, Morse theory, the
Matthai-Quillen formalism and Euler numbers of moduli spaces have also
been investigated in \blau\blauthom\ in a somewhat complementary
fashion.

\newsec{Geometry of Iterated Superspaces}

In this section we collect various mathematical facts about the
geometry of superspaces relevant to balanced topological
field theories. This can be seen as a generalization of the usual
aparatus of differential geometry, fiber bundles and cohomology
to the case of more than one differential. It will give a natural
interpretation of some of the results of \VaWi. Although one can easily
discuss the general $N_T$-extended case, we restrict ourselves 
mainly to $N_T=2$ in this section. 

\subsec{The superspace $\Xhat$}

We start with some very basic concepts.  Let us first recall that to
an $n$ dimensional bosonic manifold $X$ we can associate in a
canonical way a $(n\!\mid \!n)$ dimensional supermanifold $\Xhat = \Pi
TX$. This supermanifold is modeled on the tangent bundle $TX$ where
the parity reversion operator $\Pi$ acts by making the fibers
anti-commuting.  Any superspace is defined by its sheaf of
functions. In the case of $\Xhat$ this sheaf is generated by even and odd
coordinates $u^i,\psi^i$, where we can think of $\psi^i$ as the basis
of one-forms $du^i$. So, over an open subset $U\subset X$ the sheaf
$\CC^\infty(\Uhat)$ is given by the differential graded algebra (DGA)
\eqn\super{
\ext^*[\psi^i]\otimes \CC^\infty(U),
}
with $\ext^*$ the exterior algebra.  Analysis on the supermanifold
$\Xhat$ is equivalent to studying differential forms on $X$, {\it i.e.}\
we can
identify $\CC^\infty (\Xhat) \cong\O^*(X)$. In this way the exterior
differential $d$ is represented by the odd vector field
\eqn\diff{
d= \psi^i {\p\over \p u^i}.
}
We refer to this well-known identification that underlies much of the
applications of quantum field theory to topology as the ``supertautology.''

Before we generalize this construction to more than one differential,
we have to clarify one point. In the supergeometry of topological
field theory one often considers the differential geometry on the
total space of a vector bundle $E\to B$.  In this case we would like
to divide up the coordinates $u^i$ into two sets: ``basic
coordinates'' $u^\mu$ and ``fiber coordinates'' $\uhat^a$. Similarly,
the anticommuting variables split into $\psi^\mu=du^\mu$ and
$\psihat^a=d\uhat^a$. The structure group of the sheaf of functions on
$\widehat E$ can be reduced and it is usually convenient to use the
extra data of a connection $\nabla$ on $E$ to covariantize the action
of the differential $d$ on $\Ehat$.  Thus, our sheaf of functions will
be generated by variables $(u^\mu, \psi^\mu; \uhat^a,\psihat^a)$ with
$(\psi^\mu; \uhat^a, \psihat^a)$ transforming linearly across patch
boundaries on the base manifold $B$.  We make this a sheaf of differential
graded algebras (DGA's)
using the formula\foot{Whenever we write expressions as $R\cdot u$, we
assume that $R$ is only contracted with the linear fiber coordinate
$\uhat^a$ in $u=(u^\mu;\uhat^a)$.}
\eqn\sccnt{
\eqalign{
\nabla u & = \psi, \cr
\nabla \psi & = \half R\cdot u,\cr}
}
where $R$ is a curvature.
Here, and   subsequently, we use the obvious
identification of differential forms with polynomials in $\psi^\mu$,
{\it i.e.}\ we write $\nabla=d+\Gamma$, $\Gamma=\Gamma_\mu\psi^\mu,$
$R=\half R_{\mu\nu}\psi^\mu\psi^\nu$ etc., which act by linear
transformations on the fiber variables.  So, written out explicitly
the above equations read
\eqn\transfs{
\eqalign{
 d u^\mu & = \psi^\mu, \cr
\nabla  \uhat^a &\equiv d\uhat^a + \Gamma^a_{ \mu b} \psi^\mu \uhat^b
 =\psihat^a, \cr
 d \psi^\mu  & = 0, \cr
 \nabla \psihat^a  & \equiv
d \psihat^a + \Gamma^a_{ \mu b} \psi^\mu \psihat^b
= \half R^{a}_{b \mu \nu} \psi^\mu \psi^\nu
\uhat^b, \cr}
}
where $\Gamma^a_{ \mu b}(u)$ is the local expression for the
connection. The second line of \sccnt\ follows of course from
consistency: $\nabla^2=R$. The equations
\transfs\ should   be regarded as {\it defining} the exterior derivative
$d$ on $\Ehat$.  Summarizing,  we learn that 
in the case that $X$ is the total space of a vector bundle,
the fiber variable $\psihat^a$
should be considered as the covariant differential $\nabla\uhat^a$.

\subsec{The iterated superspace $\Xhhat$}

We now turn to the iterated superspace of $X$,
\eqn\itersuper{
\Xhhat\equiv \Pi T(\Pi T X),
}
obtained by repeating the operation of the previous section once
more.  It is also defined by its functions $F(u^i, \psi_A^i, H^i)
$. These now depend on two bosonic variables $u^i,H^i$ and two
fermionic variables $\psi^i_A$, $A=\pm$. That is, $\CC^\infty(\Xhhat)$
is the sheaf on $X$ which on an open set $U\subset X$ is the algebra
\eqn\shfi{
\ext^*[\psi_+^i,\psi_-^i]\otimes S^*[H^i] \otimes \CC^\infty(U),
}
with $S^*$ the symmetric algebra. Heuristically we think of the
variables $\psi^i_\pm$ as the one-forms $d_\pm x^i$ obtained from two
differentials $d_\pm$. The element $H^i$ can then be thought of as
$d_-d_+x^i=-d_+d_- x^i$. There are however some subtleties with a
global interpretation along these lines as we discuss in a moment.

We
obtained $\Xhhat$ by twice applying the operation $\Pi T$, but this
obscures  the natural action on the
algebra of functions of the Lie algebra $sl_2$ with generators
$J_{AB}=J_{BA}$ given by
\eqn\ntslt{
\eqalign{
J_{++} & = \psi_+^i {\p\over \p \psi_-^i}, \cr
J_{+-}& = J_{-+} = \psi_+^i {\p\over \p \psi_+^i} -
\psi_-^i {\p\over \p \psi_-^i},\cr
J_{--} & = \psi_-^i {\p\over \p \psi_+^i}. \cr} }
The operator is $J_{+-}$ is the ghost number operator that counts the
number of $\psi_+$'s minus the number of $\psi_-$'s. Under the algebra
$sl_2$ the fermions $\psi_A$ form a doublet representation, $u$ is a
singlet and $H$ a pseudo-singlet.  Here ``pseudo'' means odd under
charge conjugation $+\leftrightarrow -$. That is, the combination
$\e_{AB}H$ is invariant\foot{We use the convention that
$\epsilon^{+-} = -\epsilon_{+-} = 1$.}.
The operators $d_A, J_{AB}$ form a closed algebra:
$\{d_A,d_B\} =0$ and $d_A$ is a doublet under the $sl_2$ action.

Note that we can take an intermediate point of view and may consider
\shfi\ as defining the ring of differential forms $\O^*(\Xhat)$ on the
superspace $\Xhat$.  To do that we must break the $sl_2$ symmetry and
consider either $(\psi_+^i, H^i)$ or $(\psi_-^i,H^i)$ as one-forms.

As mentioned, we would like to turn $\CC^\infty (\Xhhat)$ into a BDGA
(bi-differential graded algebra) with differentials $d_\pm$.  How we
do this depends on how we identify the algebras \shfi\ across patches.
One approach is to identify $H^i = - d_+ \psi_-^i = d_-\psi_+^i$. This
gives a simple representation for the differentials $d_A$ as
\eqn\difff{
d_A = \psi^i_A{\p\over\p u^i} + \e_{AB} H^i{\p \over \p \psi^i_B},
}
but has the awkward feature that the variable $H^i$ does not transform
as a tensor but becomes a 2-jet with transformation rules
\eqn\twojet{
H^{i'} = {\p u^{i'}\over \p u^j}H^j + {\p^2 u^{i'}\over \p u^j \p
u^k}\psi^j_+\psi^K_-.
}
It is usually inconvenient to work with 2-jets, so we would rather
define the sheaf of functions on $\Xhhat$ by the transformation rules:
\eqn\trnsfn{
\eqalign{
\psi_A^{i'} & = {\p u^{i'}\over  \p u^j} \psi_A^j, \cr
H^{i'} & = {\p u^{i'}\over \p u^j} H^j. \cr} }
According to the discussion of section 2.1 it is then necessary to
introduce a connection $\nabla$ on the tangent bundle $TX$ and define
the exterior differentials $d_A$ in terms of this connection by the
relations
\eqn\dfnii{
\nabla_A \psi_B = \e_{AB} H,}
or, in full detail,
\eqn\transfpsi{
d_A\psi^i_B + \Gamma^i_{jk}\psi^j_A\psi^k_B= \e_{AB} H^i.
}
We will adopt this second point of view in
the development below.  It has the
consequence that the transformation of
$H$ is more complicated under
the action of the differentials $d_A$:
\eqn\htransf{
\nabla_A H = - R_{AB}\psi_C\e^{BC}.
}

The generators $U=(u,\psi_A,H)$ of $\CC^\infty(\Xhhat)$ form what we
will call a {\it basic quartet.} They can be arranged as
\eqn\genqrt{
\matrix{    &  &   \psi_+^i   &  & &\qquad\qquad\qquad 1 \cr
    & d_+ \nearrow &  &  \searrow d_- & &\cr
u^i & & & &H^i &\qquad\qquad\qquad 0 \cr
& d_- \searrow  & &\nearrow d_+ & & \cr
& & \psi_-^i & & & \qquad\qquad\qquad -1 \cr
}
}
where we have indicated the action of the differentials and the ghost
charges.  We also note as in \VaWi\ that the generators can be
conveniently combined into a $N_T=2$ superfield, by adding
two odd variables $\theta^+,\theta^-$,
\eqn\superfield{
U^i(\theta^+,\theta^-)=u^i + \theta^A \psi_A^i + \half \epsilon_{AB}
\theta^A \theta^B H^i.  }

\subsec{Vector bundles}

As we already mentioned, in many applications to topological field
theories our space $X$ actually will be the total space of a vector
bundle $E\to B$.  We will then have to consider the iterated
superspace $\itE$.  Thus, as in section 2.2, we will divide our
coordinates on $E$ again into fiber coordinates and base coordinates
$u^i=(u^\mu;\uhat^a)$. The functions on $\itE$ are now generated by
\eqn\bfrii{
U^i=(U^\mu;\Uhat^a)=(u^\mu, \psi_A^\mu, H^\mu; \uhat^a,
\psihat_A^a,\Hhat^a),
}
with $\psi_A^\mu,H^\mu; \uhat^a, \psihat_A^a, \Hhat^a$ transforming
linearly across patch boundaries.  In this case the differentials are
defined by
\eqn\brstviiii{
\eqalign{
\nabla_A u& =\psi_A,\cr
\nabla_A \psi_B & = R_{AB}\cdot u + \epsilon_{AB} H,\cr
\nabla_A H & = -R_{AB}\psi_C \epsilon^{BC}+ P_A \cdot u,\cr}
}
where the quantity $P_A$ is a three-form and $sl_2$ pseudo-doublet of
bi-degrees $(1,2)$ and $(2,1)$. It is defined by
\eqn\cpa{
P_A  = {1\over 3 } \nabla_B (R_{CA}) \epsilon^{BC}.
}
Its geometrical significance as a higher order form 
of curvature will become clear later. The appearance of terms
of this nature is one of the new features of field theories with
extended topological symmetries. For the moment we simply note that
the objects $R_{AB},P_A$ satisfy
\eqn\curvatures{
\eqalign{
R_{AB} & =\half[\nabla_A,\nabla_B],\cr
P_A  &={1\over 6} [\nabla_B,[\nabla_C,\nabla_A]]\e^{BC}.\cr
}
}
In all these formulas we use the obvious notation for identifying
bi-graded differential forms with polynomials in the fermions
$\psi_A$. That is, we have the identifications
\eqn\detail{\eqalign{
R_{AB} & = R_{\mu\nu}\psi^\mu_A\psi^\nu_B, \cr
P_A & = {1\over 3} \nabla_\mu
R_{\nu\lambda}\psi^\mu_B\psi^\nu{}_C\psi^\lambda_A \e^{BC} +
{2\over 3} R_{\mu\nu} H^\mu\psi_A^\nu.\cr}
}
Our notation is somewhat condensed. Thus, the first two lines of \brstviiii\
are shorthand for:
\eqn\fulldetail{
\eqalign{
 d_Au^\mu & =\psi^\mu_A,\cr
 d_A \uhat^a + \Gamma^a_{\mu b}\psi^\mu_A \uhat^b & =
\psihat^a_A ,\cr
 d_A\psi^\mu_B + \Gamma^\mu_{\nu\lambda}
\psi^\nu_A\psi^\lambda_B  & =
\e_{AB} H^\mu, \cr
 d_A \psihat^a_B + \Gamma^a_{~\mu b}\psi^\mu_A\psihat^b_B & =
\half R^a_{~~b\mu\nu}\psi^\mu_A\psi^\nu_B \uhat^b + \e_{AB}\Hhat^a
\cr}
}
and so on.

\subsec{De Rham cohomology of $\Xhhat$}

Given  a BDGA   one can wonder what the
properties of the corresponding cohomology theories are. A fundamental
result for what follows is the following

\medskip

{\bf Theorem 2.1}. {\it Suppose $\alpha$ is
$d_+$ and $d_-$ closed and $sl_2$ invariant, then $\alpha$ can be
decomposed as
\eqn\gnalf{
\eqalign{
\alpha & = \alpha_0 + d_+ \beta_-  + d_+ d_- \gamma \cr
& = \alpha_0 - d_- \beta_+  + d_+ d_- \gamma \cr}
}
where $\alpha_0$ is constant on components of
$X$ and $\beta_\pm= \beta_i \psi_\pm^i$, with
$\beta_i du^i \in H^1(X)$.}

\medskip

{\it Proof}.  The proof of this theorem relies on constructing
appropriate homotopy operators.  For simplicity we will work in the
case where the connection is zero (or $H$ is a 2-jet). First we note
that the algebra of functions on $\Xhhat$ is bigraded
\eqn\bigraded{
\CC^\infty(\Xhhat) = \bigoplus_{q_+,q_-\geq 0} \CC^{q_+,q_-} (\Xhhat).}
We define operators $L_\pm$ measuring the separate charges $q_\pm$
\eqn\plsch{
\eqalign{
L_\pm & = \psi^i_\pm  {\p\over \p \psi^i_\pm}  +
H^i{\p\over \p H^i }.\cr}
}
We then introduce the homotopy operators
\eqn\hmtpy{
\eqalign{
K_\pm & = \pm \psi^i_\pm {\p\over \p H^i },\cr}
}
which satisfy the algebra
\eqn\hmtpii{
\eqalign{
[d_\pm, K_\mp] & = L_\mp. \cr} }
This shows that as long as $L$ is
invertible, there is no cohomology.  Thus we learn that the $d_+$
cohomology is concentrated in degree $(q_+,0)$ while similarly $d_-
$
cohomology is concentrated in degree $(0,q_-)$.  Now, assume
$\alpha$ is
$d_A$ closed and $sl_2$ invariant and of degree $(q_+,q_-)$,
which is not $(0,0)$ or $(1,1)$.  Then we may use the homotopy
operators to show that $\alpha$ is of the form $\alpha = d_+ d_-
\gamma$ in the following way:
\eqn\dpldmn{
\eqalign{
\alpha & = {1\over  q_- (q_+-1)}
d_+ d_- (K_+ K_- \alpha)\cr
& =
{1\over  q_+ (q_- -1)}
d_- d_+ (K_- K_+ \alpha).\cr
}
}
Similarly, if $\alpha$ is of degree $(q_+,q_-)=(1,1)$ then
\eqn\spccs{
\eqalign{
\alpha & = d_-(\psi_+^i A_i(U) ) \cr
& = -d_+(\psi_-^i A_i(U)),\cr}
}
where $A_i du^i $ is a closed $1$-form on $X$. $\spadesuit$

\subsec{Vector fields and derivations}

Let $V$ be a vector field on the bosonic manifold $X$.  $V$ induces two
first order differential operators on $\Omega^*(X)$: the Lie
derivative $\CL(V)$ and the contraction $\iota(V)$ given by %
\eqn\i{
\eqalign{
\CL(V) &  = V^i {\p \over  \p u^i} + {\p V^i \over  \p u^j}  \psi^j
{\p\over \p\psi^i}, \cr
\iota(V) & = V^i{\p\over \p\psi^i} .\cr}
}
We can think of these derivations as vector fields on the superspace
$\Xhat$. In particular, $\CL(V)$ can be interpreted as the lift $\Vhat$
of the vector field $V$ to $\Xhat$.

We can now repeat this procedure and lift
the vector field \i\ to a vector field $\Vhhat$ on the iterated
superspace $\Xhhat$:
\eqn\ii{
\CL(V)=\Vhhat
=
 V^i {\p \over  \p u^i} + {\p V^i \over  \p u^j}
\psi_A^j {\p\over \p \psi_A^i}
+
\Biggl( {\p V^i \over  \p u^j} H^j - \half \epsilon^{AB}
{\p^2 V^i \over  \p u^k \p u^j}
\psi_A^j \psi_B^k \Biggr) {\p \over  \p H^i }.
}
This derivation represents the Lie derivative on functions on $\Xhhat$.

In a similar way we can represent the contraction of $\Vhat$ on forms
on $\Xhat$ to define a doublet of contractions $\iota^A(V)$ of
bi-degree $(-1,0)$ and $(0,-1)$
\eqn\iii{
\iota^A(V) =
V^i {\p\over \p \psi_A^i}
- \epsilon^{AB} {\p V^i \over \p u^j }  \psi_B^j {\p\over \p H^i}.
}
Finally, in order to close the algebra of the operators
$d_A,\iota^A,\CL$, we have to introduce the operator $I$,
a pseudo-scalar of bi-degree $(-1,-1)$ defined as
\eqn\iv{
I(V) = V^i {\p\over \p H^i}.
}
By straightforward calculation one verifies that the above operators
satisfy a bi-graded bi-differential Lie algebra based on $Vect(X)\otimes
\ext^*[\eta_+,\eta_-]$, where $\eta_\pm$ are two anticommuting
variables and the Lie bracket is given by $[V\otimes\alpha,V\otimes
\beta] = [V,W]\otimes \alpha\beta.$ One simply expands such a
derivation in its even and odd components as
\eqn\bigradedvect{
\CL(V)+ \eta_A\iota^A(V) + \half \e^{AB}\eta_A\eta_B I(V).
}
The nontrivial relations to verify are
\eqn\relations{
\eqalign{
[d_A,\iota^B(V)] & = \delta_A^B \CL(V) ,\cr
[d_A,I(V)] & = -\e_{AB}\iota^B(V) ,\cr
[d_A,\CL(V)] & = 0 ,\cr
[\iota^A(V),\iota^B(W)] & = \e^{AB} I([V,W)],\cr
[\CL(V),\iota^A(W)] & = \iota^A([V,W]),\cr
[\CL(V),I(W)] & = I([V,W]).\cr
}
}
We will make use of the operators $\iota^A, I$ when we consider
the extended equivariant cohomology in the next section.

\newsec{Extended Equivariant Cohomology}

Much of the differential geometric framework of topological field
theories is based on the concept of equivariant cohomology. Since we
are interested in models with extended topological symmetry, we will
develop in this section the notion of extended equivariant
cohomology. We will meet some interesting generalizations of the notion
of connection and curvature.

\subsec{The Weil and Cartan algebra}

In the study of the differential geometry of a principal bundle $P$
with Lie algebra $\lieg$ one encounters the so-called Weil algebra
$\CW(\lieg)$. Let us recall its definition; for more details see,
for example,
\CMRIII. The Weil algebra is a DGA with $\lieg$-valued generators
$\omega,\phi$ of degrees 1 and 2 respectively.
The action of the differential $d$
can be summarized in terms of the covariant derivative $D=d+\omega$ by
the relations
\eqn\weil{
\phi=\half [D,D],\qquad [D,\phi] =0.
}
The resulting action of $d$ is
\eqn\weill{
d\omega=\phi-\half [\omega,\omega],\qquad d\phi=-
[\omega,\phi].
}
These are of course the relations that are valid for a connection
$\omega$ and curvature $\phi$ on a principal bundle $P$.  In fact, one
can define a connection simply as a homomorphism $\CW(\lieg) \to
\Omega^*(P)$.

One interpretation of the relations \weil\ is that if we introduce the
curvature $\phi=D^2$ and take further commutators, then the process of
introducing new generators stops because of the Bianchi-Jacobi
identity $[D,[D,D]] = 0$.

One can introduce two derivations of the Weil algebra:
the interior derivative or contraction
\eqn\interior{
\iota_a\omega^b=\delta_a^b,\qquad \iota_a\phi^b=0,
}
(where $\omega=\omega^a e_a,\phi=\phi^a e_a$, with $e_a$ a basis for $\lieg$)
and the Lie derivative
\eqn\liederiv{
\CL_a=[\iota_a,d].
}
The derivations $d,\iota_a,\CL_a$ satisfy
a well known closed algebra.

The Cartan algebra $\CC(\lieg)$ is obtained by simply putting $\omega=0$
in the Weil algebra and is generated by the single variable $\phi$ of degree
two. We have the simple identity $d\phi=0$, so the action of the
differential $d$ in the Cartan algebra is completely trivial.

\subsec{Weil, BRST and Cartan models of equivariant cohomology}

There are various ways to define the equivariant cohomology $H^*_G(X)$
of a space $X$ that carries the action of a (compact) Lie group $G$.
Topologically it is defined as the cohomology of the space
$X_G=EG\times X/G$, which is the universal $X$-bundle over the
classifying space $BG$.  However, there are also algebraic
definitions.  We briefly recall the so-called Weil, BRST and Cartan
model.

For the Weil model we start with the algebra $\CW(\lieg)\otimes
\Omega^*(X)$.  On this algebra we have the action of the operators
$\iota_a$ and $\CL_a$, now defined as $\iota_a=\iota_a\otimes 1 +
1\otimes \iota_a$ {\it etc.} Here we write $\iota_a=\iota(V_a)$, with
$V_a$ the vector field on $X$ corresponding to the Lie algebra element $e_a$.
One then restricts to
the so-called basic forms $\eta\in \CW(\lieg)\otimes \Omega^*(X)$
which satisfy $\iota_a\eta=\CL_a\eta=0$. The equivariant cohomology
groups are then defined as
\eqn\equivweil{
H_G^*(X) = H^*((\CW(\lieg)\otimes\Omega^*(X))_{basic},d^W)
}
with Weil differential $d^W = d \otimes 1 + 1 \otimes d$.

The BRST model is simply related to the Weil model. One starts
from the space basic forms, but uses the differential
\eqn\brstdif{
d^{BRST}=d^W + \omega^a \otimes \CL_a - \phi^a \otimes \iota_a.
}
It was shown in \Kal\ that the two models are related as
\eqn\brstweil{
d^{BRST} = e^{\omega^a\iota_a} d^W e^{-\omega^a\iota_a}.
}

There is a simpler model of equivariant cohomology --- the Cartan
model, based on the Cartan algebra $\CC(\lieg)=S^*(\lieg^*)$. The starting
point is now the algebra $S^*(\lieg^*)\otimes \Omega^*(X)$, but as
differential we choose
\eqn\equivdiff{
d^C=1\otimes d - \phi^a \otimes \iota_a.
}
This operator satisfies $(d^C)^2= - \phi^a \otimes \CL_a$ and thus
only defines a complex on the $G$-invariant forms. The Cartan model of
equivariant cohomology is now defined as
\eqn\equivcartan{
H_G^*(X) = H^*((S^*(\lieg^*)\otimes \Omega^*(X))^G,d^C).
}
One can then show that the definitions \equivweil\ and \equivcartan\ are
equivalent and agree with the topological definition.

\subsec{Algebra of $N$-extended covariant derivatives}

If we have a geometry such as the iterated superspaces $\Xhhat$ where
it is natural to introduce several independent exterior derivatives,
then the principal bundles over such spaces will also have several
covariant derivatives. This motivates the following generalization of
the Weil algebra which we call the Weil algebra of order $N$ and
denote as $\CW^N(\lieg)$.

We introduce several covariant derivatives and connections
\eqn\covdif{
D_A=d_A+\omega_A,\qquad  A=1,\dots N,
}
and then  introduce successive ``curvatures''
\eqn\eqviii{
\phi_{A_1 \dots A_j} = {1\over  j!} [D_{A_1},[D_{A_2},\ldots,D_{A_j}]\ldots]]
}
until the Bianchi identities close the algebra.  The resulting DGA is
the Weil algebra of order $N$.  One can define various Cartan models
by putting generators to zero. We will discuss the case $N=2$ in
detail in the next subsection. Here we restrict ourselves to a general
description of the Cartan algebra $\CC^N(\lieg)$ of order $N$.

This extended Cartan algebra can be abstractly described as
follows. Let $V$ be the $N$ dimensional {\it odd} vector space
generated by the basis elements $D_A$ of degree 1. Let $L=Free(V)$ be
the {\it free} Lie algebra on $V$. This is the space spanned by all
possible commutators of the $D_A$'s imposing the relations following
from the (anti)symmetry of the Lie bracket and the Jacobi identity. So
at degree two we have the elements$[D_A,D_B]=[D_B,D_A]$ etc. Now let
$L'$ denote the subalgebra of the elements of $L$ with degree $\geq
2$. We easily see that as a Lie algebra $L'$ is generated by the
elements of $V'\equiv L'/[L',L']$, that is, $L'=Free(V')$. One can
show that $V'$ is finite dimensional and concentrated in degrees
$2,\ldots,N$. We can think of $V'$ as the space of curvatures
\eqn\phiAB{
\phi_{AB}=\half[D_A,D_B],
}
and higher order generalizations.

The elements $D_A$ in $V$ (of degree 1) act as differentials (denoted as
$d_A$) on the vector space $V'$
by
\eqn\differentials{
d_A \eta = [D_A,\eta],
}
and commute $[d_A,d_B]=0$ within $V'$. Actually, the differentials
$d_A$ will only commute up to commutator terms if we pick explicit
representatives of $V'$ in $L'$. In fact, if we pick a basis $\phi_I$
of $V'$ ($I$ will be a multi-index in terms of the indices
$A,B,\ldots=1,\ldots,N$.) and keep the same notation for its
representatives in $L'$, we have an action of $d_A$ of the form
\eqn\representatives{
d_A \phi_I = c_{AI}{}^J
\phi_J + c_{AI}{}^{JK} [\phi_J,\phi_K] + \ldots
}
where the ellipses indicate higher order commutators.
Now within $L'$ the differentials satisfy
\eqn\nolongercom{
[d_A,d_B]=2[\phi_{AB},\cdot].
}
We now define the Cartan algebra $\CC^N(\lieg)$ as the DGA
$S^*(\lieg^*\otimes V')$, where we define the differential $d_A$ by
evaluating all the Lie brackets in \representatives\ in $\lieg$. That
is, $\CC^N(\lieg)$ is the algebra generated by the $\phi_I$ which now
take values in $\lieg$.  We will see in the next section what this
all means concretely for $N=2$. Note that in contrast with the case
$N=1$, for the case $N>1$ we still have a nontrivial set of
differentials $d_A$ acting on the Cartan algebra.

\subsec{$N=2$ Weil and Cartan algebra}

We now focus on the case $N=2$, as is appropriate to $\Xhhat$. The
$N=2$ Weil model $\CW^{2}(\lieg)$ is the unique BDGA with generators
(see also \Dubois\ for a some what different definition of a bigraded
version of the Weil algebra)
\
\eqn\conncurv{
\hbox{\it connections}: \qquad
\matrix{ \omega_+ &   & \cr
  &  & \O \cr
\omega_- & & \cr}
\qquad\quad
\hbox{\it curvatures:} \qquad
\matrix{ \phi_{++} &   & \cr
  &  & \eta_+ \cr
\phi_{+-} & & \cr
 &  & \eta_- \cr
\phi_{--} & & \cr}
}
Here we indicated the ghost charges graphically. The generators
$\omega_A,\phi_{AB},\Omega,\eta_A$ have the following degrees and
$sl_2$ representations $(1,{\bf 2}),(2,{\bf 3}),(2,{\bf 1'}),(3,{\bf
2'})$. (The primes indicate pseudo-representations.)  They satisfy the
relations
\eqn\dfndee{
\eqalign{
\Omega & = d_+ \omega_- - d_- \omega_+, \cr
\phi_{AB} & = \half [D_A, D_B], \cr
\eta_A & = - {1\over 6}[D_B,[D_C,D_A]] \epsilon^{BC}. \cr}
}
We can define the Cartan model by putting $\omega_A,\O = 0$. This
leaves us just the variables $\phi_{AB}, \eta_A$ in \conncurv.  The
transformation laws become now
\eqn\crtmd{
\eqalign{
d_A\phi_{BC}& = \epsilon_{AB} \eta_C + \epsilon_{AC} \eta_B,\cr
d_A \eta_B & = -\half [\phi_{AC}, \phi_{BD}] \epsilon^{CD},\cr
}
}
reproducing the transformation laws of \VaWi.

Let us make a comment about the object $\eta_A$, because it
illustrates very well the new features of the extended
algebras. Indeed, let us see why these objects appear according to the
general definition given in the previous subsection. In the $N=2$ case
we have two covariant derivatives $D_A$ in degree one and three
curvatures $\phi_{AB}=\half[D_A,D_B]$ in degree two. In degree three
we have the triple commutators $[D_A,[D_B,D_C]]$. However, for $N=2$
the six independent triple commutators $[D_A,[D_B,D_C]]$ are not all
determined by the Jacobi identity. This should be compared to the
$N=1$ case, where the Jacobi-Bianchi identity gives us $[D,[D,D]]=[D,\phi]=0$.
In fact, for $N=2$ there are only four Jacobi-Bianchi identities which
are given by
\eqn\bianchi{
\eqalign{
[D_+, \phi_{++}] &= 0, \qquad
2[D_+,\phi_{-+}] + [D_-,\phi_{++}]  = 0,\cr
[D_-, \phi_{--}] &= 0, \qquad
2[D_-,\phi_{+-}] + [D_+,\phi_{--}]  = 0.\cr
}
}
This implies that there are two (six minus four) new generators
$\eta_A$ at degree three.  Equation \dfndee\ implies that these are
explicitly given by
\eqn\defeta{
\eta_+ = - [D_+, \phi_{+-} ] ,\qquad
\eta_- =  [D_-, \phi_{+-} ].
} One easily verifies that at degree four and higher no new generators
appear. So we learn that
\eqn\cartantwo{
\CC^2(\lieg) = S^*(\phi^a_{AB},\eta^a_A),\qquad
A,B=\pm,\; a=1,\ldots,\dim\lieg.
}

\subsec{$N=2$ extended equivariant cohomology}

We are now in a position to discuss the equivariant cohomology of
iterated superspaces.  Suppose $X$ has a $G$ action generated by
vector fields $V_a$ where $e_a$ denote a basis of the Lie algebra
$\lieg$ of $G$. By lifting these vector fields as described in section
2.5, we obtain a $G$ action on the space $\Xhhat$, together with the
derivations $d_A,\CL(V_a),\iota_A(V_a),I(V_a)$ of
$\CC^\infty(\Xhhat)$.

As in the case $N=1$, there are several models for the equivariant
cohomology, We discuss here briefly the Weil model, BRST model and
Cartan model.  For the Weil model we consider the complex
$\CW^{2}(\lieg)\otimes \CC^\infty(\Xhhat)$ and the differential is
simply the sum of the two differentials as defined above:
\eqn\weilmodel{
d_A^W = d_A\otimes 1  + 1\otimes d_A,
}
acting on the basic forms, that are now defined to satisfy
\eqn\basicforms{
\iota_A\alpha=\CL\alpha=I\alpha=0.
}

The BRST model is defined analogously as in \brstdif
\eqn\brstmdl{
d_A^{BRST} \equiv
\exp\biggl[ \iota^A(\omega_A) + I(\Omega) \biggr]
d_A^{W}
\exp\biggl[- \iota^A(\omega_A) - I(\Omega) \biggr]
}

Finally, the Cartan model is based on the $G$-invariant subalgebra  of
$\CC^2(\lieg)\otimes \CC^\infty(\Xhhat)$ with equivariant differential
\eqn\cartanmodel{
d_A^C=d_A^W + \phi^a_{AB}\iota^B(V_a) + \eta^a_A I(V_a). }
This gives the explicit transformation laws \crtmd\ together with
the following action of the Cartan differential on functions on $\Xhhat$
(or, equivalently, differential forms on $\Xhat$)
\eqn\brstvi{
\eqalign{
d_A u^i & = \psi_A^i \cr
d_A \psi_B & = \CL(\phi_{AB}) u + \epsilon_{AB} H\cr
d_A H&= -\CL(\phi_{AB}) \psi_C \epsilon^{BC} -\CL(\eta_A) \cdot
u\cr}
}
where we have dropped the superscript on $d_A$ and used
the compressed notation $\phi_{AB}=\phi_{AB}^aV_a$ {\it etc.}
This again reproduces the transformation laws  in  \VaWi.

The extended equivariant
cohomology is not very different from the
ordinary equivariant cohomology.
 One can show that the $N=2$ equivariant cohomology of $X$
is actually isomorphic to that of $X$, at least outside of degrees
$(a,b)$ for $a,b = 0,1$.  To prove this we introduce the homotopy
operator $K = K^X + K^C$ where $K^C$ for the Cartan model is defined
by $K^C \eta_+ = - \phi_{+-}, K^C
\eta_- = -\half \phi_{--}$ with $K=0$ on all other generators and
$K^X=K_-$ is defined in \hmtpy.  A short calculation shows that
\eqn\hmtpiii{
[d_+, K] = L_- + \eta^a_A {\p \over  \p \eta^a_A} +
\phi^a_{+-} {\p \over  \p \phi^a_{+-}} +
\phi^a_{--} {\p \over  \p \phi^a_{--}}
}
from which the result follows.

\newsec{Balanced Topological Field Theory}

We now introduce a new class of topological field theories, which
include the ``cofield construction'' of \CMRII\CMRIII\ as a special
case. One natural name for these theories would be $N_T=2$ topological
field theories. Here $N_T$ denotes the number of {\it topological}
supercharges or BRST operators. This should not be confused with
extended supersymmetric theories. In fact, the twisting procedure will
typically relate models with $N=2$ supersymmetry to $N_T=1$
topological symmetry and models with $N=4$ supersymmetry to $N_T=2$
topological field theories.  Since this nomenclature has perhaps too
many misleading connotations and since the ghosts and antighosts are
perfectly matched in these theories we propose to call them ``balanced
topological field theories'' (BTFT's).

\subsec{Review of the standard construction of TFT}

The basic data for a TFT are $(i)$ a space $\CC$ of fields, $(ii)$ a
bundle $E\to\CC$ of equations equipped with a metric $(,)_E$ and
connection $\nabla$ compatible with the metric, and $(iii)$ a section
$s\in \Gamma(E)$ such that its zero locus $\CM = \CZ(s)$ defines a
moduli problem of interest.  (This is reviewed in detail in \CMRIII,
see {\it e.g.}\ also \tftalso.)

The construction of the topological field theory can be phrased in
terms of the supergeometry of $\Ehat^*$. As in section 2.1 we wish to
distinguish the fiber coordinates from the field space coordinates
coordinates $u^\mu,\psi^\mu$. The fiber coordinates are the
``antighosts,'' coordinates on the dual to the bundle of equations:
\eqn\eqsns{
\eqalign{
\rho_a & \in \O^0(M;\Pi E^*),\cr
H_a  & \in \O^1(M;\Pi E^*).\cr}
}
The bundle $E^*$ carries a connection $\nabla$ with curvature $R$
and the BRST operator $Q=d$ is defined by:
\eqn\brstii{
\eqalign{
\nabla \rho & = H,\cr
\nabla H & = R\cdot \rho.\cr}
}
The topological field theory action $I$ is defined in terms of the gauge
fermion
\eqn\gaugefermion{
\Psi = i \langle \rho,s\rangle - (\rho,\nabla \rho)_{E^*}
}
in the form
\eqn\tftact{
I= Q \Psi  = i H_a s^a - (H,H)-i \langle \rho,\nabla s\rangle +
(\rho,R \rho)_{E^*}
}
where we use the compatibility of the metric and connection.  General
arguments show that the path integral $Z=\int e^{-I}$ computes the
Euler character of the bundle of antighost zero modes over the moduli
space $\CZ(s)$:
\eqn\localise{
Z = \int_{\CZ(s)} \chi(\cok \nabla s).
}

The above story becomes a little more intricate in the presence of a
gauge symmetry $G$.  The basic topological multiplet $(A,B)$ takes
values in an equivariant bundle over field space with connection
$\nabla$ and has transformation laws:
\eqn\brstiii{
\eqalign{
\nabla A & = B, \cr
\nabla B & = \CR\cdot A,\cr}
}
where the combination
$\CR  = R+ \CL(\phi) $ is the equivariant curvature \bgv.

In order to construct the Poincar\'e dual to the moduli space
$\CZ(s)/G$ one introduces the extra
 multiplet $\lambda,\eta\in \lieg=Lie(G)$ of degree $-2,-1$.
We will write here the Lie algebra indices as $\lambda^x,\eta^x$. Let
us denote the vertical vector fields associated with the gauge group
action by:
\eqn\dfnc{\eqalign{
(\CL(\lambda) u)^I &= \lambda^x V_x^I(u), \cr
V_x^I(u)& = C: \lieg \to T_u \CC\cr}
}
and define the projection gauge fermion: $\Psi_{proj} = i (\psi,
\CL(\lambda)\cdot u )$ to project out the redundant gauge degrees of
freedom.  The resulting term in the action is:
\eqn\expdi{
\eqalign{
Q \Psi_{proj}  = (\lambda ,C^\dagger & C \phi+ C^\dagger R u
+\p_J (C^\dagger)^x_I \psi^I \psi^J)  -(\psi,\CL(\eta) u). \cr}
}
Note that $\lambda$ is a Lagrange multiplier and the resulting delta
function fixes $\phi$ away from fixed points of the gauge group.

The fermion kinetic terms may be written as:
\eqn\kintrm{
i \rho_a \nabla_I s^a \psi^I + i \eta^x (C^\dagger)^x_{~I} \psi^I
=
\pmatrix{\rho & \!\eta\cr} \bO \psi,
}
where the operator $\bO$ is defined by:
\eqn\dfcplx{
 T\CC
\quad
{\buildrel \bO=\nabla s \oplus C^\dagger \over \longrightarrow}
\quad
\O^1(\CC;E) \oplus \lieg^*,
}
and is associated to the deformation complex
\eqn\dfcplx{
0\quad \rightarrow \quad
\lieg \quad {\buildrel C\over \longrightarrow}\quad  T\CC
\quad
{\buildrel \nabla s \over \longrightarrow}
\quad
E \quad \rightarrow 0 }
by using the metric.
\dfcplx\ is a complex if the equations are gauge invariant.
The complex is exact at degree $-1$, if the group action is free.
Again general arguments show that the path integral is just:
\eqn\piii{
Z = \int_{\CZ(s)/G} \chi(\cok \bO/G).
}

\subsec{Balanced topological field theories: field content}

In a balanced or $N_T=2$ topological field theory, the fields in the
model are the generators of functions on $\Xhhat$.  We will denote
coordinates on $X$ by $u^i$. Sometimes we will divide up the
coordinates into fiber and basic coordinates.  As usual the generators
form a quartet:
\eqn\genqrtii{
\matrix{    &  &   \psi_+^i  &  &  \cr
    & \nearrow &  &  \searrow & \cr
u^i &  & & & H^i\cr
 & \searrow & & \nearrow & \cr
 & & \psi_-^i & & \cr}
}
where we note that all of $\psi_A^i, H^i$ should be regarded
as (even or odd) sections of a vector bundle.
These bundles have connections so we can
define the differentials as in \brstviiii. We will
assume a group $G$ acts on $X$ and introduce
the Cartan multiplet $\phi_{AB},\eta_A$ as in \conncurv. The
$G$-equivariant BRST differentials are now defined to act by
\eqn\brstvii{
\eqalign{
\nabla_A u& =\psi_A, \cr
\nabla_A \psi_B & = \CR_{AB}\cdot u + \epsilon_{AB} H, \cr
\nabla_A H & = -\CR_{AB}\psi_C \epsilon^{BC}
+ \CP_A \cdot u,\cr}
}
where the geometrical operators are defined by
\eqn\axdfs{
\eqalign{
\CR_{AB} & = R_{AB} + \CL(\phi_{AB}),\cr
\CP_A & = {1\over 3 } \nabla_B (\CR_{CA})
\epsilon^{BC}= P_A + \CL(\eta_A),\cr
\CP_\pm & = \pm \nabla_\pm  \CR_{\pm \mp}.\cr
}
}
Here $\CR_{AB}$ and $\CP_A$ are the $N_T=2$
extended equivariant curvatures.

\subsec{Balanced topological field theories: The action potential}

A topological field theory with field space of the form
$\widehat{\widehat{\CC}}$ is called
{\it balanced} if the action is an $sl_2$ invariant and $d_+,d_-$ closed
function on $\widehat{\widehat{\CC}}$. Let us characterize the
most general action of a BTFT. The action $I$, being a
function in $\CC(\Xhhat)$, carries a bigrading $(q_+,q_-)$.
According to Theorem 2.1  the action is both $d_+$ and $d_-$
exact and is, in fact, of the form:
\eqn\actpot{
I = I_0 + d_+ d_- \CF.
}
where the ``topological term'' $I_0(u)$  is constant
on the components of $\CC$. We refer to $\CF$ as the
``action potential'' since it is analogous to the
K\"ahler potential of a K\"ahler form.
Note that $\CF$ is not uniquely defined; we can always shift
\eqn\shift{
\CF \rightarrow \CF + d_+\Phi_- + d_-\Phi_+.
}
Note further that if $H^1(\CC)\not=0$ then $\CF$ need not
be globally well-defined on field space. So the analogy to
a K\"ahler potential is quite good.

By $sl_2$ invariance,
action potentials must be of total ghost charge
zero. The most natural action potentials are of the form
\eqn\actpoti{
\CF = \CF_0(u) + \epsilon (\psi_+, \psi_-) +
\beta (H,u) + \gamma (\eta_+, \eta_-),
}
where $(\cdot ,\cdot ) $ is a metric on the bundles over
field space which is compatible with the connections
and $\CF_0(u)$ is a function on field space.
Locally, this is the most general cation potential which is at most
first order in $\psi_\pm, H, \eta_\pm$.

Let us discuss the separate terms individually.
The gauge fermions $\Psi_-=d_-\CF$ and actions
$S=d_+\Psi_-=d_+d_-\CF$ associated with these terms are:

$\bullet$ $\CF_0(u)$. We will assume that
$\CF_0(u)$ is a $G$-invariant function.
Then:
\eqn\api{
\eqalign{
\Psi_- & = \nabla_I \CF_0 \psi_-^I, \cr
d_+ d_- \CF_0(u) & = - H^I \nabla_I \CF_0
+\half \epsilon^{AB} \psi_A^I \psi_B^J
 \nabla_J \nabla_I \CF_0. \cr}
}

$\bullet$ $(\psi_+, \psi_-)$. The fermion bilinear gives rise to
\eqn\apii{
\eqalign{
\Psi_- & = (H,\psi_-) + (\CR_{-+}u,\psi_-) -(\CR_{--
}u,\psi_+), \cr
d_+ d_- (\psi_+, \psi_-) &=
- (H,H) +2 (\CP_A u, \psi_B ) \epsilon^{AB}\cr
\qquad  -  \half \epsilon^{AC} \epsilon^{BD}\biggl[  & (\CR_{AB} u,
\CR_{CD} u) +2 (\psi_A, \CR_{BC} \psi_D) \biggr] \cr}
}

$\bullet$ $(H,u)$ is equivalent to $ (\psi_+, \psi_-) $.  This follows
from the identity
\eqn\simid{
(H,u) = (\psi_-, \psi_+) - d_-(\psi_+,u).
}

$\bullet$ $ (\eta_+, \eta_-) $. This equivariant term gives the following
contributions to the gauge fermion and action
\eqn\apiii{
\eqalign{
\Psi_- & = \half ([\phi_{--},\phi_{++}],\eta_-)-
(\eta_+,[\phi_{--},\phi_{-+}]), \cr
d_+ d_-  (\eta_+, \eta_-) &=
([\phi_{++},\phi_{+-}], [\phi_{--},\phi_{-+}]) +
([\phi_{++},\phi_{--}], [\phi_{++},\phi_{--}])  \cr
& \qquad +\epsilon^{AB}\epsilon^{CD}([\eta_A,\phi_{BC}],\eta_D). \cr}
}

In section 5 below we will show that under good conditions the path
integral for the theory \actpoti\ localizes to the critical
submanifold of $\CF_0$ modulo gauge transformations:
\eqn\mdspce{
\CM = \{ u:\nabla  \CF_0(u) =0 \} / G
}
and that, moreover, the partition function computes the
Euler number of this moduli space,
\eqn\fnlans{
Z = \chi(\CM) } Thus, balanced topological field theories compute
Morse theory on field space, with the action potential serving as a
Morse function.

\subsec{Viewing BTFT as a standard TFT}

The transformation laws \brstvii\ are not standard
TFT transformations. But we may make
the redefinition $H' = \CR_{+-}u - H$ and then view the theory as a
standard one with the following familiar field content:

$\bullet$ Matter multiplets:
\eqn\stdi{
\eqalign{
\np u & = \psi_+ ,\cr
\np \psi_+ & = \CR_{++} u,\cr
\np \phi_{+-} & = -\eta_+,\cr
\np \eta_+& = -[\phi_{++}, \phi_{+-}];\cr}
}

$\bullet$ Antighosts:
\eqn\stdii{
\eqalign{
\np \psi_- & = H',\cr
\np H'  & = \CR_{++} \psi_-;\cr}
}

$\bullet$ Projection multiplet:
\eqn\stdii{
\eqalign{
\np \phi_{--} & =-2 \eta_-,\cr
\np \eta_- & =- \half [\phi_{++}, \phi_{--}]; \cr}
}

$\bullet$ Gauge fermion for equations
\eqn\gfi{
i \psi_-^I \nabla_I \CF_0 + 2 \alpha (\psi_-, \CL(\phi_{+-})u)
+ \alpha(H', \psi_-);
}

$\bullet$ Projection gauge fermion
\eqn\gfii{
-\alpha(\psi_+, \CL(\phi_{--}) u) -\half
\gamma (\eta_+, [\phi_{--}, \phi_{+-}]).
}

The rest of the gauge fermion following from
the action potential is then declared an irrelevant
$Q$-exact modification.

\subsec{The cofield construction}

The ``co-field construction'' described in \CMRII\CMRIII\VaWi\ is a map by
which we can assign a BTFT to any TFT. Under good conditions this will
compute the Euler character of the original moduli space to which the
TFT localizes.

We return to the original moduli problem in section 4.1 defined by the
vanishing of a section $s^a(u)$ in the ``bundle of equations.  The basic
idea is to take $X = E^*$ as the field space with coordinates $U^I =
(u^\mu; \uhat_a)$.  The degree $0$ part of the action potential is
then simply
\eqn\cfldi{
\CF_0(U) = \uhat_a s^a(u).
}
Clearly, the critical points of this Morse function are:
\eqn\cfldii{
s^a(u) =0, \qquad \qquad \nabla_\mu s^a \uhat_a =0.  }
If $s$ is sufficiently nondegenerate the second equation implies $\uhat_a =0$
and the solutions to the equations is the same moduli space as in the
original TFT.  By \fnlans\ we see that the BTFT will calculate the
Euler character of this moduli space.

It is straightforward to implement this idea in detail.  The fields
generate $\widehat{\widehat{E^*}}$. They may be arranged into two basic
quartets:
\eqn\fldqrt{
\matrix{    &  &   \chi^\mu   &  &  \cr
    & \nearrow &  &  \searrow & \cr
u^\mu &  & & & \Hhat^\mu \cr
 & \searrow & & \nearrow & \cr
 & & \widehat{\rho}^\mu & & \cr}
}
filling out the ``fields'' of the original
moduli problem, and
\eqn\fldqrt{
\matrix{    &  &   \widehat{\chi}_a   &  &  \cr
    & \nearrow &  &  \searrow & \cr
\uhat_a  &  & & &  H_a  \cr
 & \searrow & & \nearrow & \cr
 & & \rho_a & & \cr}
}
filling out the ``antighosts'' of the original
moduli problem.
The BTFT  action potential is:
\eqn\cfldap{
\eqalign{
\CF & = i \CF_0(U) -
\pmatrix{ \psi_-^\mu & \psi_{-,a} \cr}
\pmatrix{G_{\mu\nu} & G_\mu^{~b}\cr
G^a_{~\nu} & G^{ab} \cr}
\pmatrix{\psi_+^\nu \cr   \psi_{+,b}\cr}
\cr
\CF_0(U) & = \uhat_a s^a(u), \cr}
}
where $G$ is a metric.
The construction is easily ``equivariantized'' by
using the equivariant differentials.

\subsec{Summary: the deformation complexes}

The various classes of topological field theories
are nicely summarized by their associated deformation
complexes:

$\bullet$ For a  general {\it topological field theory}
we have the usual complex
\eqn\dfcplx{
0\quad \rightarrow \quad
\lieg \quad {\buildrel C\over \longrightarrow}\quad  T\CC
\quad
{\buildrel \nabla s \over \longrightarrow}
\quad
E \quad \rightarrow 0 }
of symmetries, fields, and equations \Witr.

$\bullet$ For a {\it balanced topological field theory}
we get the complex
\eqn\dfcplxi{
0\quad \rightarrow \quad
\lieg \quad {\buildrel (C,0)\over \longrightarrow}\quad  T\CC
\oplus \lieg
\quad
{\buildrel  (\nabla^2 \CF_0, C)  \over \longrightarrow}
\quad
T\CC \quad \rightarrow 0, }
where the maps act as
\eqn\dfcplxii{
\eqalign{
\eta_- & \rightarrow (C \eta_-, 0) \cr
& \qquad (\psi_+, \eta_+) \rightarrow \nabla^2 \CF_0 \psi_+ + C
\eta_+.\cr}
}

$\bullet$ The {\it cofield construction} is associated with the complex
\foot{The rolled-up
complex of the cofield construction suggests a role for quaternionic
vector spaces.  Moreover, these equations suggest a duality between
equations and symmetries. We thank Andrei Losev for an interesting
discussion about this. }
\eqn\dfcplxiii{
0\quad \rightarrow \quad
\lieg \quad {\buildrel  \over \longrightarrow}\quad
\bigl(E^*\oplus T\CC \bigr)
\oplus \lieg
\quad
{\buildrel  \over \longrightarrow}
\quad
\bigl(E^*\oplus T\CC \bigr) \quad \rightarrow 0, }
where the maps are defined as
\eqn\dfcplxii{
\eqalign{
\eta_- & \rightarrow (C \eta_-, 0, 0) \cr
& \qquad (\psihat_+^a, \psi_+^i , \eta_+) \rightarrow
\nabla^2 \CF_0 \psi_+ + C \eta_+.\cr}
}

\newsec{Localization of BTFT}

In this section we justify the localization result \mdspce\ and \fnlans\
more fully.  As we have seen, the general action potential
can be taken to be a sum of a function $\CF_0$ on field space
and quadratic terms in the fermions $\psi_A$ and $\eta_A$:
\eqn\apv{
\CF = i \CF_0(u) + \alpha(\psi_+, \psi_-)
+ \gamma (\eta_+,\eta_-),
}
where $\alpha, \gamma$ are constants.
Putting $\alpha$  to zero results in a singular Lagrangian and an
ill-defined path integral. The coefficient $\gamma$ is subtle and is
related to the introduction of mass terms into topological field theory.
The general discussion of the localization of the theory based on the action
potential \apv\ is quite involved. We will simply illustrate it for
the following situation:

$(i)$ All the curvatures and connections can be set to zero. This is
the case for topological Yang-Mills, where $\CC$ is an affine space
and for 2D topological gravity in the Beltrami formulation.

$(ii)$ The gauge group $G$ acts without fixed points.

$(iii)$ The coefficient $\gamma=0$. (Otherwise the action is
not quadratic in $\phi_{++}, \phi_{--}$. )

$(iv)$ All the zero modes of the Hessian of $\CF_0$ on critical
submanifolds are associated with gauge symmetries or tangent
directions to the moduli space.

In the case that the conditions $(i)$--$(iv)$ are satisfied, we can
justify the localization to \mdspce\ above, as we will now
demonstrate. Let us introduce the notation
\eqn\notts{
(\psi_1, \CL(\phi)\cdot \psi_2)_{T^*\CC}
\equiv (\phi, K(\psi_1,\psi_2))_\lieg\qquad .
}
 The action \apv\ becomes:
\eqn\actionsum{
d_+ d_- \CF = L_1 + L_2+ \alpha L_3 + \alpha L_4,
}
with
\eqn\smpact{
\eqalign{
L_1 & = -i \langle \nabla \CF_0, H\rangle -
\alpha (H,H), \cr
L_2 &  =  \psi_+^I (\nabla^2 \CF_0)_{IJ} \psi_-^J
+ 2 (\psi_+, C \eta_-) - 2 (\psi_-, C \eta_+), \cr
L_3 &=  (\phi_{+-}, C^\dagger C \phi_{+-})-2(\phi_{+-},
K(\psi_-,\psi_+)) ,\cr
L_4 & =  -(\phi_{--}, C^\dagger C \phi_{++}-K(\psi_+,\psi_+))
+ (\phi_{++}, K(\psi_-,\psi_-) ) .\cr}
}

The four terms of the Lagrangian play distinguished roles in the
evaluation of the path integral, and can be discussed separately:

$\bullet$ $L_1$ is the familiar localization to the critical points of
the action potential. The evaluation of the path integral near these
critical points gives
\eqn\dti{
{1\over  \mid \Det' \nabla^2 \CF_0\mid }.
}

$\bullet$ $L_2$ is the fermion Lagrangian associated with the
deformation complex
\dfcplxi\ and \dfcplxii\ of the equations
$\nabla \CF_0 =0$.  Note that gauge invariance of $\CF_0$ guarantees
that this is a complex since $\nabla^2\CF C \eta=0$ at the critical
points.

Note that the virtual dimension of the moduli space is automatically
zero: in the balanced theory there are as many ghost zero modes as
antighost zero modes, and they live in the same bundle.  The fermion
operator is thus:
\eqn\frmop{
\pmatrix{ \nabla^2 \CF_0 & C \cr C^\dagger & 0 \cr}.
}
Because we assume that $\CF_0$ is a nondegenerate
Morse function we can block diagonalize into
the kernel of $\nabla^2 \CF_0$ and its orthogonal
subspace:
\eqn\frmopi{
\pmatrix{
(\nabla^2 \CF_0)' & 0 & 0 \cr 0 & 0 & C \cr 0 & C^\dagger & 0 \cr}
}
There is also a finite-dimensional space of fermion zero modes
associated to the tangent to moduli space, or, better,  to the
cohomology of the complex \dfcplxi.

The determinant of the fermion non-zero modes is
\eqn\detii{
\det' (\nabla^2 \CF_0) \cdot \det (C^\dagger C).
}

$\bullet$ $L_3$: The integral is gaussian, so that $\phi_{+-}$
effectively localizes to zero.  (More precisely, it localizes to an
even nilpotent.)  The path integral gives:
\eqn\detiii{
{1\over \sqrt{ \det C^\dagger C} }
\exp {1 \over  \alpha}\Biggl[ (K(\psi_-,\psi_+), {1\over  C^\dagger
C}
K(\psi_-,\psi_+) ) \Biggr].
}

$\bullet$ $L_4$: This is also a gaussian integral and gives:
\eqn\detiv{
{1\over \det C^\dagger C }
\exp{1 \over  \alpha}
\Biggl[ (K(\psi_+,\psi_+), {1\over  C^\dagger C}
K(\psi_-,\psi_-) ) \Biggr].
}

Notice that the determinants of $C$'s do not cancel.  The reason is
that we have not fixed the gauge.  This can be very elegantly solved
using the differential topology that we introduced in section 3.  We
can include naturally the ghosts {\it as well as the antighosts} of
$G$-gauge fixing by passing to the Weil model (instead of the Cartan
model) of equivariant cohomology, and introducing a
gauge-noninvariant term in the action potential.

Recall that in the case $N_T=2$ the Weil multiplet consists of a
triplet $(\omega_+,\omega_-,\Omega)$ of connections, see \conncurv.
Here the connection $\omega_+$ appears as the ghost. The connections
$(\omega_-,\Omega)$ represent the antighost multiplet.  The gauge
fixing Lagrangian is written as
\eqn\gfxing{
\eqalign{
d_+ d_-(\epsilon u^2) & =
\epsilon d_+(\omega_-, C^\dagger u)  + \epsilon d_+(u, \psi_-) \cr
& = \epsilon (\Omega, C^\dagger u) + \epsilon (\omega_-, C^\dagger C
\omega_+) + \epsilon d_+(u, \psi_-). \cr} }
The integrals over the
first two terms provide the missing $ \sqrt{ \det C^\dagger C}$. The
last term adds some gauge-noninvariant pieces to the ``matter''
Lagrangian, but we can invoke $\epsilon$-independence to argue that
these terms make no contribution.

The net result of the path integral is an integral
over collective coordinates:
\eqn\ntrslt{
\int_\CM \prod d u_0^I  d \psi_+^{0,I} d \psi_-^{0,I}
\exp\Bigl((K(\psi_+,\psi_+), {1\over  C^\dagger C}
K(\psi_-,\psi_-) \Bigr).
}

Finally, let us recall that if $E_{1,2}$ are trivial hermitian
vector bundles and $A$ is a linear fiber map
$A: E_2 \to E_1$ then there is a natural connection
on
$\ker A^\dagger\subset E_1$
given by $P\circ d$ where $P$ is the
projection operator. The curvature is just
\eqn\gncnsd{
R = P dA {1 \over  A^\dagger A} d A^\dagger P
}
In our case $C: \lieg \to T\CC$ and the tangent
bundle to the moduli space is $T\CM\cong \ker C^\dagger$.
We recognize this form in the remaining integral \ntrslt.
Putting all this together we obtain the result
\fnlans.

\subsec{Localization of the cofield model: ``counting without signs''}

The cofield model can be put into the standard framework by taking the
field space to be $E^*\to M$ and the antighost bundle to be
$\pi^*(E\oplus T^*M) \to E^*$.  We choose the section $ {\bf s} = (s ,
\nabla_\mu s^a \uhat_a) $ and localize to \cfldii.  For simplicity
suppose $\nabla_\mu s^a$ has no kernel and the index is all
cokernel. Then we localize to $\uhat =0$.  Note that the fermionic
operator is
\eqn\cflhss{
\nabla^2 \CF_0 = \pmatrix{ 0 & \nabla_\mu s^a \cr
(\nabla_\mu s^a)^\dagger & 0 \cr} }
For this reason the fermionic path
integral is always positive semidefinite and we are ``counting without
signs'' \VaWi.  In any case, the result is: $Z=\chi(\CZ(s))$, which
was, of course, the original motivation for the cofield construction
\CMRII.

\newsec{Examples of BTFT's}

In this section we briefly mention some important examples of balanced
topological field theories in various dimensions. Note that in principle
we have a map that
 associates to any local QFT action $\CF$ a BTFT, by simply
using $\CF$ as action potential. Of course, to get a reasonable
action for the BTFT, for example quadratic in derivatives, the action potential
should satisfy certain constraints. Typically it will
be first order in derivatives. Fortunately, there are quite a few
interesting candidates of that form.

\subsec{Morse theory}

Take $X$ to be a finite dimensional Riemannian manifold and $\CF_0$ to
be a Morse function. This is the standard example to which
supersymmetric quantum mechanics on $X$ ($SMQ(X)$) reduces.
The path integral becomes:
\eqn\zrdm{
Z = \int \exp\Biggl[-i H^\mu \nabla_\mu \CF_0 -
\epsilon G_{\mu\nu}H^\mu H^\nu +
\epsilon[ (\psi_+, \nabla^2 \CF_0 \psi_-) + \cdots
\Biggr]
}
where the ellipses indicate various curvature
terms.
Note that if
$
\CF_0 = \half \sum \lambda_i u_i^2
$
then the quadratic term in the Lagrangian is
$(\nabla \CF_0)^2 = \sum \lambda_i^2 u_i^2$. This is indeed
the canonical example. If we choose
$U=(u,\psi_\pm,H)\in \widehat{\widehat{\bR}}{}^n= \bR^{2n\mid2n}$
with $\CF(U)={i\over 2}uAu + \psi_+B\psi_-$ and
$A,B$ quadratic forms, the fundamental gaussian integral is
\eqn\gaussian{
{1\over (2\pi i)^n} \int_{\widehat{\widehat{\bR}}{}^n}
\exp d_+d_- \CF = sign(\det A).
}
The determinants cancel, and the result does not depend on the choice
of $A$ and $B$, up to a sign. So we see that $Z$ reduces to the
sum of the indices of the critical points, and indeed equals the
Euler number $\chi(X)$.

\subsec{Balanced quantum mechanics}

Ironically, one cannot obtain $SQM(X)$ as a balanced theory, in spite
of the fact that $Z=\chi(X)$ for $SQM(X)$ \trminus.
The balanced theory must necessarily have an action of the form:
\eqn\bqm{
\int_{S^1} dt\, \omega_{\mu\nu} [ \dot x^\mu H^\nu
- \dot \psi_+^\mu \psi_-^\nu  - \epsilon
H^\mu H^\nu],
}
where $\omega= \omega_{\mu\nu} dx^\mu dx^\nu$ is
a closed two-form.

A very natural class of such theories is provided by a symplectic
target space $(X,\omega)$. Our field space is in that case $LX$, the space of
closed unbased loops.  The action potential leading to \bqm\ is just
\eqn\bqmi{
\CF_0  = \oint_{S^1} \alpha_\mu \dot x^\mu
= \int_D x^* \omega,
}
where $d \alpha = \omega$ and we consider the
circle to be the boundary of a disk $D$.  Moreover, if $H(x(t),t)$ is
a time-dependent Hamiltonian then it is natural to consider the more
general action potentials:
\eqn\bqmii{
\CF_0  = \int_D x^* \omega
+ \oint  H(x(t),t) dt
}
Morse theory based on this functional is
the subject of symplectic Floer homology
\floer.

\subsec{Balanced $\sigma$-models}

There are many natural action potentials one might want to consider in
the context of sigma-models.  For example $\CF_0 = \int (\nabla f)^2 $
would lead to a theory which calculates the Euler character of the
moduli space of harmonic maps. Closely related
actions have appeared in
\polyakov\horava.  Other obvious choices are the Nambu action
$\cF_0=Area(f(\Sigma))$.  Such actions lead to nonrenormalizable
actions. For example, the harmonic map choice leads to an action
fourth-order in derivatives. (N.B. The theory is easily generalized to
four dimensions). For this reason we focus on a particular case,
described in the next section.

\subsec{Cofield $\sigma$-models}

We describe the cofield construction for topological sigma models
\CMRII\CMRIII. Begin with the standard moduli problem from
holomorphic maps: $E \to MAP(\Sigma, X)$.
The fields in $\widehat{\widehat E}$ fit into two quartets:
\eqn\csmiii{
\matrix{    &  &   \psi^i   &  &  \cr
    & \nearrow &  &  \searrow & \cr
x^i &  & & & \bar H^i \cr
 & \searrow & & \nearrow & \cr
 & &\bar\pi^i & & \cr}
\qquad\quad
\matrix{    &  &   \pi^{\zb}_{~i}    &  &  \cr
    & \nearrow &  &  \searrow & \cr
p^{\zb}_{~i}  &  & & &  \bar H^{\zb}_{~i}  \cr
 & \searrow & & \nearrow & \cr
 & &\bar \psi^{\zb}_{~i}  & & \cr}
}
and
\eqn\csmiiv{
\matrix{    &  &   \bar \psi^\ib   &  &  \cr
    & \nearrow &  &  \searrow & \cr
\bar x^\ib &  & & & H^\ib \cr
 & \searrow & & \nearrow & \cr
 & & \pi^\ib & & \cr}
\qquad\quad
\matrix{    &  &   \bar \pi^z_{~\ib}    &  &  \cr
    & \nearrow &  &  \searrow & \cr
\bar p^z_{~\ib}  &  & & &  H^z_{~\ib} \cr
 & \searrow & & \nearrow & \cr
 & & \psi^z_{~\ib}  & & \cr}
}
where $i,\bar i$ are holomorphic (anti-holomorphic) indices on the
target space $X$ and $z,\zb$ are holomorphic (anti-holomorphic) on the
worldsheet.  As we will discuss in section 7 these fields will
describe a conformal field theory.

The action potential is:
\eqn\csmv{
\CF^{BT\sigma} = i \CF_0 +\CF^{metric},
}
where
\eqn\bvbvbv{
\eqalign{ \CF_0 & = \int_\Sigma \sqrt{h} \biggl[ p^{\zb}_{~i}  \pb_{\zb} x^i
+ \bar p^z_{~\ib}  \p_z \bar x^\ib \biggr],\cr
\CF^{metric} & =
\int_\Sigma
 \sqrt{h}  h^{z \zb}\Biggl[
\pmatrix{\bar \psi^{\zb}_{~i}  & - \bar\pi^i  \cr} {\bf L}
\pmatrix{ \bar \pi^z_{~\jb} \cr \psib^\jb}
 +
\pmatrix{ \psi^{z}_{~\jb}  &-  \pi^\jb  \cr} {\bf L}^{Tr}
\pmatrix{  \pi^\zb_{~i } \cr \psi^i } \Biggr], \cr}
}
where $h$ is a metric on $\Sigma$ and
  ${\bf L}$ is a metric related to the hyperk\"ahler metric on $T^*X$, as
described in section 7 below.

If there is a moduli space but  no antighost zero modes, {\it i.e.}\ if
$\dim \cok D_\zb = 0 , \dim \ker D_\zb >0$  where
$D_\zb=\pb_{T^*X} : \Omega^{0,0}(\Sigma;T^*X)\to
\Omega^{0,1}(\Sigma;T^*X)$,
then we localize to $\hat f^\alpha_\mu=0$, $f\in HOL(\Sigma,X)$,
the space of holomorphic maps. Furthermore, the path integral is
given by $Z= \chi(HOL(\Sigma,X))$. This situation is
uncommon.

\subsec{Balanced topological 2D gravity: Beltrami
formulation}

There is also a balanced version of topological gravity. We
can give two (equivalent) definitions, either using the
language of metrics or of complex curves. We start with the latter
point of view. In that case the relevant moduli problem is a pair
$(C,V)$ with $C$ a complex curve of genus $g$ and $V$ a holomorphic
vector field on $C$. Since for $g>1$ such a vector field is generically zero,
the moduli space reduces to the moduli space $\CM_g$ of curves.
However, the virtual dimension of the moduli problem
is zero and the theory is thus balanced.
So, by definition this model computes $\chi(\CM_g)$.

In more detail: We fix a complex structure and consider the
Beltrami differentials $\mu_\zb^z\in \CB^{(-1,1)}$ which modify
the Dolbeault operator to
\eqn\newdblt{
\pb^{(\mu)} = \pb_{\zb} + \mu \p_z.
}
The deformation complex becomes:
\eqn\bltrplxi{
0\quad \longrightarrow \quad
Vect^{1,0} \quad {\buildrel C\over \longrightarrow}\quad  \CB^{(-
1,1)}
\oplus Vect^{1,0}
\quad
{\buildrel D  \over \longrightarrow}
\quad
\CB^{(-1,1)} \quad \longrightarrow \quad 0, }
with
\eqn\ops{
C\eta_- =\pmatrix{ \pb^{(\mu)} \eta_- \cr [\hat f, \eta_-] \cr},
\qquad
D (\mu, \eta_+) = ([\hat f, \mu] , \pb^{(\mu)} \eta_+),
}
where the ``cofield'' $\hat f$ is a vector field, and one must take care to
write:
\eqn\vfld{
\eqalign{
(\pb^{(\mu)} V)^z {\p \over \p z} &  \equiv
\Biggl[ ( \pb_{\zb} + \mu \p_z) V^z + [\mu,V]^z)
\Biggr] {\p \over \p z}, \cr
[\mu,V]^z & = \mu \p_z V^z - V^z \p_z \mu.\cr}
}

As we mentioned above, interpreted as an ordinary
topological field theory
we have the moduli problem of a holomorphic vector
field and a complex curve $(C,V)$. For $g>1$ there
are no nonsingular holomorphic vector fields.
Thus, the localization to $\hat f = 0 $ makes sense
and the path integral computes the orbifold
Euler character of moduli space.

\subsec{Balanced topological 2D gravity: metric formulation}

An alternative formulation of balanced topological gravity
starts from metrics. Let $MET$ denote
the space of Riemannian metrics $h_{\alpha\beta}$
on a topological surface $\Sigma$.
The basic quartet of $\widehat{\widehat {MET}}$ will be denoted as
\eqn\grvqrt{
\matrix{    &  &  \psi_{\alpha \beta,+}  &  &  \cr
    & \nearrow &  &  \searrow & \cr
h_{\alpha \beta} &  & & &k_{\alpha \beta}  \cr
 & \searrow & & \nearrow & \cr
 & &  \psi_{\alpha \beta,-} & & \cr}
}
We continue to take  $\Diff(\Sigma)$ as the gauge group, since there are no
Weyl-invariant metrics on $MET$. A natural choice of
action is
\eqn\balgrav{
I =  d_+ d_-(\int \sqrt{h} h^{\alpha \beta} k_{\alpha
\beta}),
}
but an equivalent and more convenient choice of action potential is:
\eqn\grvi{
\eqalign{
\CF^{BTG} &= (\psi_+,\psi_-),\cr
d_- \CF^{BTG} & = (H,\psi_-) + (\CR_{-+}u,\psi_-) -(\CR_{--
}u,\psi_+) \cr
& = \int \sqrt{h} \Biggl[\rho^z_\zb(D_z  
\hat f^\zb -H_z^\zb) + c.c.
\Biggr]
 + \int \sqrt{h} \lambda^\alpha \nabla^\beta \psi_{\alpha \beta}.
\cr}
}
Translating the fields
to the standard notation for 2D gravity (see, e.g. \CMRIII,
sec. 16.2)
we have:
\eqn\jhjhj{
u_{\alpha \beta} \to \delta h_{\alpha \beta}, \qquad
\psi_- \to \rho,\qquad
\phi_{+-} \to \hat f^\alpha, \qquad
\phi_{--} \to \lambda^\alpha,
\qquad \phi_{++}\to \gamma^\alpha.
}
Again
we recognize the gauge fermion appropriate to
the moduli problem of a pair $(C,V)$, $C$ a
curve and $V$ a holomorphic vector field.
\foot{
We still must choose an action potential that
fixes the Weyl mode. In principle any\
Diff$(\Sigma)$-invariant functional
of the metric $\CF_0[h_{\alpha\beta}]$
which takes a  {\it unique}  minimum in
each conformal class can serve as $\CF_0$. }

\subsec{ Balanced topological strings}

The coupling of the sigma model to gravity is simply  summarized by
taking the sum of the action potentials
$\CF= \CF^{BTG} + \CF^{BT\sigma} $ and using the\ Diff-equivariant version of
$d_A$. This completely encodes the coupling to
balanced topological gravity ! Let us study the resulting
coupling to gravity.

We separate the BRST operator into the part that
varies the graviton and the rest:
\eqn\grvvr{
d_A = d_A^0 + \psi_{A,\alpha \beta} {\delta \over  \delta
g_{\alpha\beta}}.
}
The action may then be expressed as
\eqn\cpgrv{
\eqalign{
d_+ d_- \CF = d_+^0 d_-^0  \CF
& + \epsilon^{AB} d_A^0
\Biggl({\delta \CF \over
\delta g_{\alpha\beta} }\Biggr)
\psi_{B, \alpha \beta} \cr
+
\Biggl({\delta \CF \over
\delta g_{\alpha\beta} }\Biggr)
(k_{\alpha \beta} + \cdots)\; +\; &
\Biggl({\delta^2 \CF \over
\delta g_{\alpha\beta} \delta g_{\gamma \delta} }\Biggr)
\psi_{+,\alpha\beta}\psi_{-,\gamma \delta} \cr}
}
Thus, the auxiliary field $k_{\alpha\beta}$ couples
to the stress tensor $K_{\alpha \beta}$ of the action potential,
while the two partners $\psi_{A,\alpha\beta}$ of the
graviton couple to the variations of the gauge fermions:
\eqn\spcrrnt{
G_{A,\alpha\beta} \equiv d_A^0
\Biggl({\delta \CF \over
\delta g_{\alpha\beta} }\Biggr)
= {\delta \Psi_A^0 \over
\delta g_{\alpha\beta}.}
}
The four currents $K_{\alpha\beta}, G_{A,\alpha\beta}, T_{\alpha\beta}$
fit into a quartet:
\eqn\crrqrt{
\matrix{    &  & G_{+,\alpha\beta} &  &  \cr
    & \nearrow &  &  \searrow & \cr
 K_{\alpha\beta} &  & & & T_{\alpha\beta}\cr
 & \searrow & & \nearrow & \cr
 & &G_{-,\alpha\beta}& & \cr}
}
Specifically, for the cofield sigma model:
$K_{zz} = \pi_{zi} \p x^i + \ldots, K_{\zb\zb}= \bar \pi_{\zb\ib}
\pb_{\zb} x^\ib + \ldots$

\subsec{2D Yang-Mills}

There is an obvious choice for an action potential that is
first order in derivatives for a
two-dimensional gauge theories. Consider a connection $A$ together
with a Lie-algebra valued scalar field $\phi$ on a Riemann surface
$\Sigma$. Choose the action potential 
\eqn\yangmills{
\CF = \int {\rm Tr}(\phi F),
}
which has its critical points on the moduli space of flat
connections on $\Sigma$. The resulting action will be of the
form $I=\int F_{\mu\nu}^2 + \cdots$. In fact, this model is rather familiar,
since it corresponds directly to the reduction to two dimensions of
four-dimensional Donaldson theory \Witdgtr.

\subsec{3D Chern-Simons}

For a three-dimensional gauge theory there is also a
canonical choice for a first-order action potential: the famous
Chern-Simons term.  Note that quite generally, for any
gauge theory we have the field quartet
\eqn\ymqrt{
\matrix{    &  &   \psi_{+,\mu}  &  &  \cr
    & \nearrow &  &  \searrow & \cr
A_\mu &  & & & H_\mu\cr
 & \searrow & & \nearrow & \cr
 & & \psi_{-,\mu}& & \cr}
}
while the Cartan multiplet $\phi_{AB}, \eta_A$
are $\lieg$-valued fields on spacetime $X$.

If we choose the three-dimensional action potential
\eqn\ymiii{
\CF = \int_X \Tr ( A d A + {2\over  3} A^3 ) + \Tr\, \psi_+ \psi_-,
}
the resulting action is, according to our general
formulae:
\eqn\ymiv{
\eqalign{
I = & \int_X \Tr\biggl[ FH - \epsilon H^2 + 2(D \eta_+ \psi_- - D \eta_-
\psi_+)
\cr & \ \ \ \ + \psi_+ [ \phi_{--} ,\psi_+] + \psi_- [ \phi_{++}
,\psi_-] -2 \psi_-[\phi_{+-}, \psi_+] \cr &\ \ \qquad + (D
\phi_{+-})^2 - D \phi_{++} D \phi_{--} \biggr] \cr}
}
 This turns out to be the reduction to three dimensions of Donaldson
theory. The Morse theory problem in this case defines Floer's
3-manifold homology theory. The theory computes the Euler number of
the moduli space of flat connections on $X$.  One can similarly
discuss the $IG$ theories of
\wttngrv.  See also the work of \blau.

\subsec{4D Yang-Mills}

This is the context of the twisting of $N=4$ supersymmetric Yang-Mills
discussed in \VaWi.  We now consider the cofield construction applied
to Donaldson theory. This should calculate the Euler character of the
moduli space of self-dual instantons.  According to the cofield
construction we should have two quartets:
\eqn\ymvi{{\rm fields}\quad
\matrix{    &  &   \psi_{+,\mu}  &  &  \cr
    & \nearrow &  &  \searrow & \cr
A_\mu &  & & & H_\mu\cr
 & \searrow & & \nearrow & \cr
 & & \psi_{-,\mu}& & \cr}
\quad {\rm equations:}\quad
\matrix{    &  &   \psi_{+,\mu\nu}    &  &  \cr
    & \nearrow &  &  \searrow & \cr
B_{\mu\nu}  &  & & & H_{\mu\nu}   \cr
 & \searrow & & \nearrow & \cr
 & &  \psi_{-,\mu\nu} & & \cr}
}
In addition we have the Cartan quintet for YM gauge symmetry, as above.

The naive cofield construction would suggest the action potential
$\CF_0 = \int_X B F^+$, but to match with the twisted action of
$N=4$ SYM one must take a modified action potential. The correct choice is
\eqn\ymviii{
\eqalign{
\CF & = \CF_1 + \CF_2 + \CF_3 \cr
\CF_1 & = \int_X \Tr(
 B^{\mu\nu} F_{\mu\nu}^+
+ {1\over  12} B^{\mu\nu} [B_{\mu\lambda},B^\lambda_\nu] ) \cr
\CF_2 & =
 \int_X \Tr( \psi_-^{\mu \nu} \psi_{+,\mu\nu}
+ \psi_-^\mu \psi_{+,\mu}  ) \cr
\CF_3 & = \int_X \Tr( \eta_+ \eta_-) \cr}
}

The balanced 4D YM theory may be identified with a twist of
the $N=4$ SYM theory  as described in \VaWi.
We embed $SU(2)_R$ into the internal $SU(4)$ symmetry of $N=4$
SYM so that $2+2 = 4$. The fermion multiplets then become:
\eqn\twsfi{
\eqalign{
(2,1,4) \oplus (1,2,\bar 4)
& = (2,2) \oplus (2,2) \oplus (1,1) \oplus (1,3)\oplus (1,1)\oplus
(1,3)\cr
&
= (\psi_{+,\mu} ) \oplus (\psi_{-,\mu} )  \oplus (\eta_+) \oplus
(\psi_{+,\mu\nu} )\oplus (\eta_-)\oplus (\psi_{-\mu\nu}) \cr}
}
The unbroken internal $SU(2)$ symmetry is
the $sl_2$ symmetry of the balanced theory.
Similarly we obtain the scalars from
$\psi^{IJ} = (4\times 4)_{antisymm} \rightarrow 3 \times (1,1)
 + (1,3) $ giving
$B_{\mu\nu}  ,
\phi_{AB}  $.
Adding $\CF_3$ makes the twisted theory
closer to the physical theory, by giving a
potential energy to the scalars.

\newsec{A new twist on the topological sigma model}

Just the way the balanced four-dimensional Yang-Mills theory
is related by a twist to the $N=4$ supersymmetric Yang-Mills
theory, the balanced topological string
on a K\"ahler target space $X$ is closely
related to an $N=4$ string with target
space $T^*X$. These strings are
quite interesting, since the balancing
property implies that they are critical
in any dimension.\foot{The work
in this section was done in collaboration with K. Intriligator
and R. Plesser.}

\subsec{Free $N=4$ Multiplets}

As a warmup we consider a single free
$N=4$ multiplet that we write as
\eqn\v{\eqalign{
X_{A\Bd} & = \pmatrix{ x &\bar p \cr - p & \xb \cr} \cr
\psi_{A\Bd}& = \pmatrix{ \psi & \pib \cr - \pi  & \psib \cr}
\cr}
}
When we generalize to $d$ such multiplets we
should regard it as defining a hyperk\"ahler
sigma model with target space $T^* \IC^d$.
On shell, this theory  has
 a large $N=4$ superconformal symmetry. We focus on
the aspects that generalize to a more general
target $T^*X$ with $X$ a K\"ahler manifold.
The four supercurrents are
\eqn\nfrcrr{
G_{\Ad \Bd} = \psi_{A\Ad} \p X_{B \Bd} \epsilon^{AB}
=\pmatrix{ \pi \p x - \psi \p p & \pi \p \bar p + \psi \p \xb\cr
- \psib\p x - \pib \p p & - \psib\p \bar p + \pib \p \bar x \cr}
}
We furthermore have an $SU(2)_L \times SU(2)_R$ current
algebra. The right currents are given by
\eqn\rtcrr{
\eqalign{
J_{\Ad \Bd} & = \half \psi_{A\Ad} \psi_{B\Bd} \epsilon^{AB}.\cr}
}
These three currents correspond to the three K\"ahler forms
$\omega_C, \omega_R, \omega_C^*$ in the case of a general hyperk\"ahler
manifold.

We will be interested in targets for which there is an additional
$U(1)$ isometry of the metric. In the present
case the  $U(1)$ isometry current is:
\eqn\ismcrr{
\eqalign{
J^{isom}_z & = \pi_{+,i}  \pib_{+,\ib}  - p_i \p \bar p_{\ib} \cr
\tilde J^{isom}_\zb & = \bar\pi_{-,\ib}  \pi_{-,i}  - \bar p_\ib \pb
p_{i }\cr
}
}
Note that the isometry current is not a conformal current, and
the conservation law is $\pb J_z - \p \tilde J_{\zb} =0 $.
Nevertheless, if one proceeds naively and
evaluates the OPE's for on-shell fields, one finds
that the charges of the fields under this current are:
\eqn\chrges{
\eqalign{
J^{isom}_z(z) \cdot \pi(w) & \sim  {1\over  z-w} \pi(w) \cr
J^{isom}_z(z) \cdot  p(w) & \sim  {1\over  z-w} p(w) \cr
J^{isom}_z(z) \cdot \p \bar p(w) & \sim  -{1\over  z-w} \p \bar p(w) \cr}
}
Now recall that the standard topological twist of an $N=4$ multiplet
is defined as the following modification of the stress-tensor
\eqn\stdtwst{
\eqalign{
T'  & = T + \p J_{\dot + \dot -} \cr
\tilde T' & = T - \pb \tilde J_{\dot + \dot -} \cr}
}
This gives the standard A-model for $T^*\IC^d$ \wittenmirror.

We now describe the new twist, which we
call the ``isometry twist'' or ``I-twist.''
In terms of conformal field theory
the isometry twisted model is related to
the $T^*\IC^d$ A-model by the twists:
\eqn\stdtwst{
\eqalign{
T'' & = T + \p J_{\dot + \dot -} - \p J^{isom} = T' -  \p J^{isom}\cr
\tilde T'' & = \tilde T'- \pb \tilde J^{isom}\cr}
}

The field content of the I-twisted model
off shell is described by the bosonic fields
$x^i, \bar x^\ib, p^\zb_i , p^z_{\ib} $, the
ghost number one fields $\psi^i , \psib^\ib, \pi^\zb_i, \pib^z_{\ib} $,
and the  ghost number $-1$ fields
$\psib_z^\ib , \psi_\zb^i , \pi_i, \pib_{\ib} $.
On shell, we have holomorphic fields:
$p_{zi} , \bar p^z_\ib, \psi^i, \psib_z^\ib, \pi_{zi} , \pib_{\ib}$
and similarly for anti-holomorphic fields.
In particular, the anti-holomorphic bosonic fields
include
$\tilde p_{\zb \ib} , \tilde{\bar p}^{\zb}_i,
\tilde{\bar \pi}_{\zb \ib}, \tilde \pi_i$.
We summarize a comparison
 of dimensions for holomorphic
conformal fields in the following table. Here $\Delta'$ and $\Delta''$
indicate the conformal dimensions in the usual A-twist and the new
I-twist respectively.

\begintable
 operator | $\Delta'$  | $\Delta''$ \elt
 $\psi$ | 0 | 0  \elt
 $\bar \psi$ | 1 | 1  \elt
 $\pi$ | 0 | 1  \elt
 $\pib$ | 1 | 0  \elt
 $p$ | 0 | 1  \elt
 $\p \bar p$ | 1 | 0  \elt
$ \pi \p x - \psi \p p$ | 1 | 2\elt
$  \pi \p \bar p + \psi \p \xb$ | 1 | 1\elt
$- \psib\p x - \pib \p p$ | 2 | 2\elt
$- \psib\p \bar p + \pib \p \bar x$ | 2 | 1\elt
$J_{\dot +\dot +}$ | 0 | 1\elt
$J_{\dot +\dot -}$ | 1 | 1\elt
$J_{\dot +\dot +}$ | 2 | 1
\endtable

Note the unusual feature that in the I-twist a bosonic
current gets twisted. This is one of the most interesting aspects of the
isometry twist. Note also that the isometry current is BRST exact:
\eqn\exact{
J^{isom} = \{ \oint  \pi \p \bar p + \psi \p \xb, - p \pib\}.
}
Thus, even though it is not a good conformal current, the resulting model is
well-defined.

The currents from the isometry twist couple to
gravity as in \crrqrt\ with:
\eqn\actptcp{
\eqalign{
K & = p \p x,\cr
G_+ & = \pi \p x + p \p \psi,\cr
G_- & = \psib \p x - p \p \pib ,\cr
T & = \p \bar x \p x + p \p (\p \bar p) + \pi \p \pib +
\psib \p \psi. \cr}
}

\subsec{Hyperk\"ahler metric on $T^*X$}

Suppose $X$ is a K\"ahler manifold with metric $G_{i \jb}d x^i d
x^{\jb}$ and corresponding K\"ahler form $\omega$.  Let $K_0$ be the
K\"ahler potential.  The noncompact manifold $T^*X$ has
a hyperk\"ahler metric (of signature $(n,n)$) $\bG$ on $T^*X$
\calabi. To make this plausible
note that $c_1(T^*X)=0$ and that, in terms of local holomorphic
coordinates $(z^i,p_i)$ on $T^* X$, there is a very natural
nonvanishing holomorphic 2-form: $\omega_C = dz^i \wedge dp_i$.
We denote the components of this hyperk\"ahler metric on $T^*X$ as:
\eqn\hypk{
ds^2 =  \bG_{i \jb} dx^i d\bar x^\jb
+  \bG^i_{ \jb}  Dp_i  d\bar x^\jb+
 \bG_{i}^{  \jb} dx^i D\bar p_\jb+  \bG^{ i   \jb} Dp_i  D\bar p_\jb
}
The K\"ahler potential is of the form
\eqn\clbi{
K=f(\xi),\qquad  \xi=G^{i\jb}(x) p_i \bar p_{\jb} = \parallel p
\parallel^2
}
and hence the metric has the required
$U(1)$ isometry in the tangent directions.

{\it Example}. One example of this construction
has appeared in the theory of the $N=2$ string
\ogvf.  Let $X$ be the upper half plane with Poincare metric
and $\xi = (Im z)^2 \parallel\! w \! \parallel^2$.
Then the K\"ahler potential for the hyperk\"ahler metric is:
\foot{Actually, the signature is $(2,2)$ so the
metric is {\it hypersymplectic}.
See \barret\
for a careful discussion of the
signs involved. }
\eqn\exctmt{
K=2 \sqrt{c \xi + e^2} + e \, \log\biggl[  {\sqrt{c \xi + e^2} -e\over
\sqrt{c \xi + e^2} + e}\biggr]
}
To avoid a singularity we must
take $c>0$.  The construction is $SL(2,\IR)$ invariant
and thus defines a hyperk\"ahler metric on the cotangent
bundles to Riemann surfaces of genus $g>1$.

\subsec{Isometry twisted $\sigma$-model in the general case}

Using the isometry we twist in the manner described above.  In order
to do this with the sigma model action one must first add a
topological K\"ahler term to make the bosonic part of the action chiral.
The momentum coordinates pick up conformal spins $\pm 1$.
Consequently the off-diagonal parts of the metric $\bG$ obtain
conformal spin.

We then proceed as follows. Define
\eqn\hypmet{
\bG = \pmatrix{ h^{z\zb} \bG_{i \jb}  &  \bG^{zi}_{ \jb}\cr
 \bG_{i}^{ \zb \jb} & \bG^{ i   \jb}\cr}
}
The gauge fermion of the I-twisted model will
be:
\eqn\itwstgf{
\Psi_-= \int\sqrt{h}
\Biggl[ \pmatrix{ \psib_z^\jb & \pib_\jb \cr} \bG
\pmatrix{\pb x^i - H_\zb^i \cr D_\zb p^\zb_i - H_i\cr}
+
\pmatrix{ \psi_\zb^i &  \pi_i  \cr} \bG^{tr}
\pmatrix{\p x^\jb - H_z^\jb \cr D_z p^z_\jb - H_\jb\cr}
\Biggr]
}
To relate this model to the balanced $\sigma$ model we must
relate the fields. We take:
\eqn\rltflds{
\eqalign{
 \pmatrix{ \psi^{z}_{~\jb}  \cr -  \pi^\jb  \cr} & =
   \bG \pmatrix{ \psi_\zb^i \cr  \pi_i  \cr}  \cr
 \pmatrix{ H^{z}_{~\jb}  \cr  H^\jb  \cr}
& =
   \bG \pmatrix{ H_\zb^i \cr  H_i  \cr}
\cr
\pmatrix{\bar \psi^{\zb}_{~i}  \cr - \bar\pi^i  \cr}
& =
\bG^{tr} \pmatrix{ \psib_z^\jb \cr \pib_\jb \cr}
\cr}
}
and ${\bf L}= \bG^{-1}$.

In order to relate this theory to the actual action
written in \CMRII\CMRIII\  we need to use that
for $\xi\to 0$ the  hyperk\"ahler potential has the form:
\eqn\limk{
K\to K_0 + a \xi +\CO(\xi^2) +
F(z^i) + F(z^i)^*
}
so that near $\xi=0$ the metric becomes a product metric. Since
the theory localizes to $\xi=0$ (thus effectively killing
half the bosonic degrees of freedom) the theories are effectively
the same.

\subsec{Relation to the $N=2$ String}

The matter systems described above appear in the $N=2$ string. Indeed,
the twisting of the $N=4$ theory was used in \brkvf\ to produce topological
field theory formulae for certain $N=2$ string amplitudes. However,
the gravitational sector of the $N=2$ string and the balanced
topological string appears to be different.

The string theory of large $N$ 2D Yang-Mills theory is a balanced
topological string, and that lead to a conjecture that the balanced
topological string for 4D balanced topological string is related to
the large $N$ limit of the Donaldson invariants
\mooreicm\CMRIII.  A slightly different conjecture has been
put forward in \ovii\ relating the $N=2$ string to the large $N$ limit
of ``holomorphic Yang-Mills''  \park.
A better understanding of relation of the gravitational
sector of balanced topological gravity and the gravitational sector of the
$N=2$ string might shed some light on the compatibility of these
conjectures, and even on the nature of 4D topological gauge theories.

\newsec{Concluding Remarks}

Some aspects of the above discussion deserve further
investigation. For example, the isometry twist provides a novel method
of eliminating bosonic zero modes, and thus provides a novel means of
dimensional reduction. Also, there are subtle issues related to
the fact that the current used in the twist is {\it not} a conformal
current.

Naively, the absence of interesting cohomology on $\Xhhat$ suggests
that there are no interesting observables. Moreover, the cancellation
of the anomaly reinforces this. However, this is probably too naive
since the action is itself $d_+ d_-$ exact and yet the path integral
is not zero.  This point remains to be clarified.  The fact that
balanced topological strings exist in any dimension is quite
curious. In view of this it would be exciting to introduce observables
into the theory.

Recently there has been intense study of
``Dirichlet branes''  or D-branes
\ref\chjp{S. Chaudhuri,
C.Johnson and J. Polchinski, ``Notes on
D-branes,'' hep-th/9602052.}.
BPS states associated
to D-branes are counted by Euler characters of
certain moduli spaces. It would be interesting to
see if one can apply BTFT's to the study of
D-branes.

\bigskip
\centerline{\bf Acknowledgements}

We thank K. Intrilligator and R. Plesser for some initial
collaboration on this project.  We also thank A. Losev, N. Nekrasov,
J.-S. Park, S. Shatashvili, I. Singer, G. Thompson and G. Zuckerman
for useful discussions.  R.D. likes to thank the Yale Physics
Department for hospitality during part of this work.  This work was
supported by DOE grants DE-FG02-92ER40704, DOE-91ER40618, by NSF grant
PHY 91-23780 and by a Presidential Young Investigator Award.

\listrefs

\bye